\documentclass{aa}  

\usepackage{natbib}
\bibpunct{(}{)}{;}{a}{}{,} 
\usepackage{graphicx}
\usepackage{txfonts}
\usepackage[colorlinks=true,linkcolor=blue,citecolor=blue, urlcolor=blue]{hyperref}
\usepackage{xcolor}

\newcommand{\teff}{$T_{\rm eff}$}
\newcommand{\eexc}{$E_{\rm exc}$}
\def\vt{$\xi_{\rm t}$}
\def\kms{$\rm km~s^{-1}$}
\def\ione{\,{\sc i}}
\def\ii{\,{\sc ii}}

\newcommand{\eps}{\log\varepsilon}
\newcommand{\gkai}[1]{}

\begin{document}

\title{
Unlocking the mystery of Sr synthesis in the early Galaxy through analysis of barium isotopes in very metal-poor stars.
\thanks{
Based on observations collected at the European Southern Observatory Science Archive and the Keck Observatory Archive.}}
\titlerunning{Unlocking the mystery of Sr synthesis}
\authorrunning{Sitnova et al.}

\author{T. M. Sitnova\inst{1}, L.~Lombardo\inst{2}, L. I. Mashonkina\inst{1}, F. Rizzuti\inst{3,4,5}, G. Cescutti\inst{6,4,5}, C. J. Hansen\inst{2}, P. Bonifacio\inst{7}, E.~Caffau\inst{7}, A. Koch-Hansen\inst{8}, G. Meynet\inst{9}, R. Fernandes de Melo\inst{2}}

\institute{
Institute of Astronomy, Russian Academy of Sciences, Pyatnitskaya 48, 119017, Moscow, Russia, \email{sitamih@gmail.com}\
\and
Goethe University Frankfurt, Institute for Applied Physics (IAP), Max-von-Laue-Str. 12, 60438, Frankfurt am Main, Germany
\and
Heidelberger Institut f{\"u}r Theoretische Studien, Schloss-Wolfsbrunnenweg 35, D-69118 Heidelberg, Germany
\and
INAF, Osservatorio Astronomico di Trieste, via Tiepolo 11, I-34143 Trieste, Italy
\and
INFN, Sezione di Trieste, via Valerio 2, I-34134 Trieste, Italy
\and
Dipartimento di Fisica, Sezione di Astronomia, Università di Trieste, Via G. B. Tiepolo 11, 34143 Trieste, Italy
\and
LIRA, Observatoire de Paris, Universit{\'e} PSL, Sorbonne Universit{\'e}, Universit{\'e} Paris Cit{\'e}, CY Cergy Paris Universit{\'e}, CNRS, 92190 Meudon, France
\and
Zentrum f\"ur Astronomie der Universit\"at Heidelberg, Astronomisches Rechen-Institut, M\"onchhofstr. 12, 69120 Heidelberg, Germany
\and
Department of Astronomy, University of Geneva, Chemin Pegasi 51, 1290 Versoix, Switzerland
           }

\date{Received ; accepted }

\abstract
{}
{
We determine the contributions of the rapid (r) and slow (s) neutron capture processes to the Ba isotope mixture, along with Ba, Eu, and Sr NLTE abundances, in a sample of very metal-poor stars. The selected stars formed before the contribution from the main s-process in low- and intermediate-mass stars became significant. Some of our sample stars are enhanced in Sr, with [Sr/Ba] reaching up to 0.7. These stars gained their high Sr abundance from a poorly understood process, sometimes referred to in the literature as a light element primary process, which may appear to be a weak s-process or a weak r-process. Our aim is to uncover the nature of this additional Sr source.
}
{
The abundances derived from the resonance Ba\ii\ 4554 and 4934~\AA\ lines are influenced by the adopted Ba isotope mixture. We compute Ba isotope mixtures corresponding to different r- to s-process contributions (pure r-process, 80\%/20\%, 50\%/50\% and 12\%/88\%, i. e. solar ratio) and determine the corresponding abundances from the Ba\ii\ resonance lines in each sample star. Additionally, we determine Ba abundances from weak subordinate Ba\ii\ lines, which are unaffected by the adopted Ba isotope mixture. We then compare the Ba abundances derived from the subordinate lines with those from the Ba\ii\ resonance lines.
}
{
We find a higher s-process contribution to Ba isotopes in stars with greater [Sr/Eu] and [Sr/Ba] overabundances, suggesting that the additional Sr synthesis was due to the early s-process occurring in massive stars. Using Sr-enhanced stars, we estimate the [Sr/Ba] ratio produced by the early s-process and obtain [Sr/Ba]$_{\rm earlyS}$ = 1.1 $\pm$ 0.2. 
The derived value should be regarded as an upper limit, as we cannot definitively exclude the possibility of a contribution to Sr from the weak r-process, which produces Sr but not Ba. Regarding the potential synthesis of Sr and Ba in the i-process in massive stars, our results for Ba isotopes and element abundances argues that there was no detectable contribution from this process within the error bars in our sample stars.
}
{
In the early Galaxy, before significant main s-process enrichment, barium and strontium were produced primarily by the main r-process and the early s-process, which occurred in rapidly rotating massive stars.
}

\keywords{Galaxy: halo --  Stars: abundances}

\maketitle

\section{Introduction}

The synthesis of light neutron-capture (n-capture) elements (Sr, Y, Zr) has been puzzling astrophysicists for several decades. Detailed observational constraints on the origin of heavy elements and the early Galactic chemical evolution can be derived from spectroscopic studies of very metal-poor (VMP, [Fe/H]\footnote{We use a standard designation, [X/Y] = $\log($N$_{\rm X}$/N$_{\rm Y}$)$_{*} - \log($N$_{\rm X}$/N$_{\rm Y}$)$_{\odot}$, where N$_{\rm X}$ and N$_{\rm Y}$ are total number densities of elements X and Y, respectively.} $< -2$) stars, which, as the oldest objects, provide a unique opportunity to study the earliest epoch of the Galaxy formation. 

Literature data reported a 2~dex spread in their [Sr/Ba] ratio, ranging from [Sr/Ba] = $-0.4$ in stars where n-capture elements originate purely from the main rapid (r) n-capture process \citep{2003ApJ...591..936S,2005A&A...439..129B,2009A&A...504..511H,2010A&A...516A..46M,2018ApJ...865..129R,2024MNRAS.529.1917S}, up to a high [Sr/Ba] of 1.5~dex \citep[see, for example, pioneering studies and some more recent works by][and many others studies]{1978A&A....67...23S,1994A&A...287..927G,1995AJ....109.2757M,2001A&A...376..232M,2005ApJ...632..611A,2007A&A...476..935F,2008ARA&A..46..241S,2013AJ....145...26R,2012A&A...545A..31H,2014ApJ...797..123H}. The latter suggests the presence of an additional source of light n-capture elements beyond the main r-process.
The main slow (s) n-capture process does not fit this role, as it occurs in asymptotic giant branch (AGB) stars with masses from 1 to 6~M$_{\odot}$, and its contribution to the Galactic chemical evolution becomes significant after a delay of at least 1 Gyr, detectable in stars with [Fe/H] $> -2$ \citep{2004ApJ...601..864T,2004ApJ...617.1091S,2011RvMP...83..157K,2020ApJ...900..179K}.

The need for an additional source of light n-capture elements operating in the early epoch of Galactic chemical evolution was first postulated by \citet{2004ApJ...601..864T} and it was named the light element primary process (LEPP). However, the LEPP was introduced as a phenomenological process without any specific hypotheses regarding its astrophysical site or nucleosynthesis mechanism.

The nature and nucleosynthesis mechanism of this extra source of the light trans-iron elements remain debated. However, it is generally believed to be linked to massive stars, as they evolved rapidly and started to enrich the interstellar gas at the earliest phases of Galactic chemical evolution. Several hypotheses have been proposed for the source associated with the s-process: the weak s-process in massive stars during the hydrostatic core He-burning phase \citep{1991ApJ...367..228R} and the shell carbon-burning phase \citep{1991ApJ...371..665R}; the s-process in fast-rotating, extremely metal-poor (EMP, [Fe/H] $\leq -3$) massive stars \citep{2008ApJ...687L..95P,2011Natur.472..454C,2013A&A...553A..51C,2014A&A...565A..51C,2021MNRAS.502.2495R} occurring in the same phases. For observers, a key difference between the above predictions is that models incorporating stellar rotation predict the production of the second-peak s-process elements (Ba, Ce, La, etc.), whereas standard weak s-process models stop at the first-peak s-process elements (Sr, Y, Zr). The s-process models in massive stars have been explored across a wide range of masses, metallicities, and rotational velocities \citep{2012A&A...538L...2F,2016MNRAS.456.1803F,2018A&A...618A.133C,2018ApJS..237...13L}. A key signature of the s-process is a production of even Ba isotopes $^{134}$Ba and $^{136}$Ba, which are s-process only isotopes. Consequently, the chemical evolution model of \citet{2013A&A...553A..51C} and \citet{2014A&A...565A..51C} predicts that VMP stars with high [Sr/Ba] should exhibit an enhanced value of the even/odd isotopes. This results from their formation from s-process material produced by rapidly rotating massive stars.

Other proposed scenarios include: charged-particle reactions in the high-entropy wind of young neutron stars \citep{1992ApJ...395..202W,2007PhR...442..237Q}; the weak r-process in core collapse supernovae \citep{2009ApJ...692.1517I,2014JPhG...41d4005A}; the $\nu$p-process in  proton-rich supernovae ejecta \citep{2006PhRvL..96n2502F,2018JPhG45a4001E,2022ApJ...929...43G}; the intermediate n-capture process \citep[i-process, ][]{1977ApJ...212..149C} in low-metallicity AGB stars with masses below 4~M$_{\odot}$ \citep{2021A&A...648A.119C,2024A&A...684A.206C}, and the i-process in massive VMP stars \citep{2018ApJ...865..120B}. These processes, except for the i-process, produce Sr but not Ba, while the i-process  yields both Sr and Ba, with a predominance of odd barium isotopes $^{135}$Ba and $^{137}$Ba \citep{2024A&A...684A.206C}.  However,  the i-process struggles to explain VMP stars with high [Sr/Ba]: the first scenario \citep{2024A&A...684A.206C} is attributed to low- and intermediate-mass stars, which enrich the interstellar gas with a delay, while the second one \citep{2018ApJ...865..120B} does not efficiently produce Sr, predicting close to or below solar [Sr/Ba] ratios.

To better understand the source of additional Sr, we aim to answer the following questions:
(i) Is the source of extra Sr the s-process or r-process?
(ii) Does this source produce a significant amount of Ba?
(iii) If so, which Ba isotopes are formed?

Our key diagnostics for identifying the n-capture element sources are: (i) [Ba/Eu] as a proxy for r-process contribution; (ii) [Sr/Ba] as a tracer of the extra Sr production source in the earliest epoch; (iii) the Ba isotope ratio as a key indicator distinguishing between s- and r-process.

The paper is structured as follows. In Sect.~\ref{sample_obs}, we describe our sample stars and observations. The abundance determination method is presented in Sect.~\ref{abund}. A discussion of our findings is provided in Sect.~\ref{discussion}, and our conclusions are summarised in Sect.~\ref{conclusions}.

\section{Stellar sample and observations}\label{sample_obs}

Our sample consists of 16 VMP giant stars from the Chemical Evolution of R-process Elements in Stars (CERES) survey \citep[][hereafter LB22]{2022A&A...665A..10L}. The selected stars have normal carbon abundances \citep{2024A&A...691A.220F}, and their chemical composition is typical to the Milky Way (MW) halo stars with the corresponding metallicity. The majority of the sample stars fall within a metallicity range of $-3.10 <$ [Fe/H] $< -2.45$, with one star having [Fe/H] = $-2.15$. Regarding the origin of the n-capture elements in the sample stars, they either have [Ba/H] $< -2.4$ \citep[][hereafter LH25]{2025A&A...693A.293L} and formed before the contribution from the main s-process in low- and intermediate-mass stars became significant  \citep[][LH25]{2004ApJ...617.1091S}, or they are r-process enhanced stars with $-2.4 <$ [Ba/H] $< -2.2$. Three of the sample stars are strongly r-process-enhanced r-\ii\ type stars with [Eu/Fe]\footnote{Here we employ a notation of \citet{2004A&A...428.1027C} for r-process enhanced stars classification.} = 1.3, while seven stars are r-\ione\ type, with 0.3 $<$ [Eu/Fe] $<$ 0.8. Thus, the selected stars trace the earliest epoch of Galactic chemical evolution and n-capture element synthesis.

For Ba isotope ratio analysis, we selected stars with high-resolution observed spectra covering both Ba\ii\ resonance lines at 4554~\AA\ and 4934~\AA. Additionally, the equivalent widths (EWs) of these lines should not exceed 140 m\AA. Sixteen of the fifty-two CERES stars meet these criteria, with their Ba\ii\ resonance line EWs ranging from 42~m\AA\ to 135~m\AA. The selected stars, along with their stellar atmosphere parameters, are listed in Table~\ref{table_par}. Our selection criteria provide a suitable stellar sample, where the Ba\ii\ resonance lines are strong enough to trace Ba isotope ratios while remain unsaturated enough to allow for accurate abundance measurements.

\begin{table*}
\caption{Stellar sample, atmospheric parameters, and characteristics of the observed spectra used for the Ba\ii\ 4554 \AA\ line analysis.} 
\setlength{\tabcolsep}{1.0mm}            
\label{table_par}      
\centering          
\begin{tabular}{l l l l l l l l }   
\hline      
\tiny{Name$_{\rm SIMBAD}$}  &  \tiny{Name$_{\rm CERES}$} & \teff$^1$ ,  & log g & \tiny{[Fe/H]}        & \vt , &  \tiny{S/T$^2$} & Program ID; PI \\
                            &                           &  \tiny{K} & \tiny{$\rm cm~s^{-2}$} &     & \tiny{\kms} &           &         \\
\hline     
HD4306           & CES0045--0932 & 5020 & 2.29 & --2.95 & 1.60 & H & C36H; S. Castro\\  
HD13979          & CES0215--2554 & 5080 & 2.00 & --2.73 & 1.90 & H & U09H; R. Kraft\\  
HD27928          & CES0422--3715 & 5100 & 2.46 & --2.45 & 1.60 & H & C23H; I. Ivans\\
HD107752         & CES1222+1136  & 4830 & 1.72 & --2.91 & 1.75 & H & C04H; J. Cohen\\ 
HD108317         & CES1226+0518  & 5340 & 2.84 & --2.38 & 1.50 & U & 165.N-0276(A); R. Cayrel \\
HD122563         & CES1402+0941  & 4680 & 1.35 & --2.79 & 2.00 & U & 266.D-5655(A); S. Bagnulo \\
HD126587         & CES1427--2214 & 4910 & 1.99 & --3.05 & 1.60 & H & U65H; M. Bolte \\
HD128279         & CES1436--2906 & 5280 & 3.15 & --2.15 & 1.30 & U & 71.B-0529(A); D. Silva \\
HE0336--2412     & CES0338--2402 & 5240 & 2.78 & --2.81 & 1.50 & H & C2230; M. Brown\\
HE1320--1339     & CES1322--1355 & 4960 & 1.81 & --2.93 & 1.90 & U & 170.D-0010(G); N. Christlieb\\ 
HE2229--4153     & CES2232--4138 & 5190 & 2.76 & --2.58 & 1.50 & U & 170.D-0010(G); N. Christlieb \\
HE2327--5642     & CES2330--5626 & 5030 & 2.31 & --3.10 & 1.55 & U & 170.D-0010(G); N. Christlieb \\ 
BD+23~3130       & CES1732+2344  & 5370 & 2.82 & --2.57 & 1.50 & H & U080Hb; J. Prochaska\\
BPS~CS31078--018 & CES0301+0616  & 5220 & 3.01 & --2.93 & 1.30 & H & U10H; M. Bolte\\
BPS~CS29491--069 & CES2231--3238 & 5220 & 2.67 & --2.77 & 1.50 & U & 170.D-0010(G); N. Christlieb\\
BPS~CS22186--023 & CES0419--3651 & 5090 & 2.29 & --2.81 & 1.60 & U & 68.B-0320(A); N. Christlieb\\ 
\hline      
\multicolumn{8}{l}{1 -- The effective temperatures from LB22 have been rounded to the nearest 10~K.}\\ 
\multicolumn{8}{l}{2 -- Spectrograph/Telescope: U -- UVES/UT2, H -- HIRES/Keck\ione.}\\ 
\end{tabular}
\end{table*}

For our sample stars, high-resolution spectra were obtained by LB22 with the UV-Visual Echelle Spectrograph (UVES) at the UT2 Kueyen Telescope, with high signal-to-noise (S/N) ratios $> 150$ per pixel at 3900~\AA. Details of the individual LB22 spectra, including S/N ratio, spectral resolution, exposure time, observation date, and settings, are provided in Table A.1 of LB22. The LB22 spectra cover a wide wavelength range and allow us to analyse the lines of Ba\ii\ at 5853~\AA, 6141~\AA, 6494~\AA, and 4934~\AA, but not the Ba\ii\ 4554~\AA\ line, due to a wavelength gap between 4520~\AA\ and 4780~\AA. Therefore, in this study, we complement the observations from LB22 with additional spectra covering the Ba\ii\ 4554~\AA\ line, obtained either with UVES or the High-Resolution Echelle Spectrometer (HIRES) at the Keck~\ione\ Telescope. The adopted spectra are of high-quality, with a spectral resolution of $\lambda/\Delta \lambda >$ 40~000 and a signal to noise ratio of S/N $>$ 100. These spectra are available from the European Southern Observatory Science Archive and the Keck Observatory Archive, respectively. The proposal IDs and PIs for the spectra used in the Ba\ii\ 4554~\AA\ analysis are listed in Table~\ref{table_par}.

\section{Abundance analysis}\label{abund}
\subsection{Codes and model atmospheres}
In this study, we use classical 1D model atmospheres from the \textsc{marcs} model grid \citep{marcs}, interpolated for the given \teff, log~g, and [Fe/H] of the stars, while \textsc{atlas12} model atmospheres \citep{2005MSAIS...8...14K} were adopted in the CERES project (LB22 and LH25). We have verified that using either \textsc{marcs} or \textsc{atlas12} models yields consistent results.

We solve the coupled radiative transfer and statistical equilibrium equations using the \textsc{detail} code \citep{Giddings81,Butler84}, incorporating the updated opacity package presented by \citet{mash_fe}. For synthetic spectra calculations, we use the \textsc{synthV\_NLTE} code \citep{Tsymbal2018}, attached to the \textsc{idl binmag} code \citep{2018ascl.soft05015K}. This method allows us to obtain the best fit to the observed line profiles while accounting for the non-local thermodynamic equilibium (NLTE) effects via pre-calculated departure coefficients (the ratio between NLTE and LTE atomic level populations) for a given model atmosphere. When fitting the line profiles, the abundance of the element of interest is varied alongside the macroturbulent velocity (v$_{\rm mac}$) and the radial velocity (v$_{\rm r}$).

The spectral synthesis line list is extracted from a recent version of the Vienna Atomic Line Database \citep[VALD,][]{2019ARep...63.1010P,2015PhyS...90e4005R}, which provides isotopic and hyperfine structure components for a number of studied chemical elements. We adopt the oscillator strengths recommended by VALD.

\subsection{Atmospheric parameters and impact of their uncertainties on abundance determination}

The selected sample stars are VMP giants with parallaxes ranging from 0.17 to 7.63 mas \citep{2021A&A...649A...1G}. Photometric effective temperatures (\teff) and distance-based surface gravities (log~g) are adopted from LB22. The uncertainty in \teff\ is 100~K, and the uncertainty in log~g amounts to 0.08~dex. In this study, we focus on the analysis of Ba\ii\ lines and the comparison of barium abundances derived from the resonance (strong) and subordinate (weak) lines.

The uncertainties in \teff\ and log~g affect abundances from different Ba\ii\ lines in the same way and do not significantly impact the abundance difference between the resonance and subordinate lines of Ba\ii. For example, in a model atmosphere with \teff/log~g/[Fe/H] = 5100/2.46/$-2.5$, a shift in \teff\ of 100~K results in an abundance shift of 0.10 and 0.07~dex for the Ba\ii\ 4554~\AA\ and 5853~\AA\ lines, respectively, with EWs of 120~m\AA\ and 25~m\AA. Overall, this shift in \teff\ produces a minor effect of 0.03~dex on the abundance difference between the two lines. In the same model atmosphere, a change in log~g of 0.08~dex results in a similar abundance shift of 0.03~dex for both lines and does not affect the abundance difference.

Regarding variations in the microturbulent velocity (\vt), an uncertainty of 0.2~\kms\ has a significant effect on the abundance derived from the resonance line, causing a shift of 0.10~dex, while the corresponding shift for the Ba\ii\ 5853\AA\ line is negligible. In our sample stars, the EWs of the Ba\ii\ resonance lines vary from 60 to 130~m\AA. For the weaker Ba\ii\ resonance line, with EW$_{\rm 4554}$ of 60~m\AA, an uncertainty in \vt\ of 0.2~\kms\ results in an abundance shift of 0.04~dex.

\subsection{Determination of \vt }
LB22 relies on \vt\ calculated using an empirical formula provided in \citet[][hereafter MJ17]{2017A&A...604A.129M}. The adopted formula was derived from \vt\ determinations based on Fe\ione\ and Fe\ii\ lines in a sample of VMP giants and provides an accuracy of 0.2~\kms. In our study, the uncertainty in \vt\ is crucial, since it mostly contributes to an uncertainty in barium abundance from the resonance lines. We refine \vt\ using Ti\ii\ lines and adopt \vt\ values that yield consistent titanium abundances from Ti\ii\ lines with different EWs.

Unlike the conventional use of iron lines for \vt\ determinations, we select Ti\ii\ lines because they have accurate oscillator strengths based on the laboratory measurements of \citet{Wood2013_ti2}, exhibit minimal NLTE effects, and are convenient to use. The strongest lines of Ti\ii\  are of comparable strength to the Ba\ii\ resonance lines in the corresponding model atmospheres. For example, in the case of the Ba\ii\ 4554~\AA\ line, the equivalent widths are EW$_{\rm 4554}$ = 60~m\AA\ and 124~m\AA, for HE0336--2412 and HD27928, respectively. For the Ti\ii\ 4468~\AA\ line, the equivalent widths are EW$_{\rm 4417}$ = 85~m\AA\ and 109~m\AA\ in the same two stars, respectively.

Although the NLTE effects for Ti\ii\ are not large \citep{2011MNRAS.413.2184B,sitnova_ti,2022A&A...668A.103M}, they are not negligible. We account for deviations from LTE using the Ti\ione-\ii\ model atom from \citet{sitnova_ti}, updated with data for inelastic collisions with hydrogen atoms (Ti\ii\ $+$ H) based on quantum-mechanical calculations presented in \citet{2020AstL...46..120S}. A grid of NLTE abundance corrections (the difference between NLTE and LTE abundance) for Ti\ii\ lines covering our stellar parameter range is provided in Table~\ref{grid_tab}. For each sample star, we select the microturbulent velocity that provides consistent abundances from weak and strong Ti\ii\ lines, with abundance slopes as a function of EW of $\le 10^{-4}$ dex per m\AA.

\begin{table}
   \caption{NLTE abundance corrections and equivalent widths in m\AA\ for Ti\ii\ lines as a function of \teff, log~g, and [Fe/H].}
   \label{grid_tab}
 \setlength{\tabcolsep}{1mm}
   \begin{tabular}{rrrrrrr}
      \hline
\multicolumn{7}{l}{$\lambda$, \AA\  species  \eexc, eV  log gf } \\
T$_{\rm eff 1}$, K & log~g$_1$ & log~g$_2$ &  log~g$_3$ & log~g$_4$ & log~g$_5$ & log~g$_6$ \\
$\rm [Fe/H]_1$ & \small{EW$_1$} & \small{EW$_2$} &  \small{EW$_3$} & \small{EW$_4$} & \small{EW$_5$} & \small{EW$_6$} \\
$\rm [Fe/H]_1$ & $\Delta_1$ & $\Delta_2$ & $\Delta_3$ & $\Delta_4$ & $\Delta_5$ & $\Delta_6$  \\
      \hline
\multicolumn{7}{l}{4417.710,  Ti\ii\ , \eexc\ =  1.16,  log gf = --1.19} \\
 \teff = 5000 &  1.0 & 1.5 & 2.0 & 2.5 & 3.0 &   3.5  \\
   --2.0 &   -1 &    124 &    106 &     92 &     79 &     --1\\
   --2.0 &   -1 & --0.094 & --0.085 & --0.065 & --0.049 &     --1\\
      \hline
   \end{tabular}\\
The table is accessible in a machine-readable format at the CDS. A portion is shown to illustrate its format and content. If EW = $-1$ and $\Delta_{\rm NLTE} = -1$, this means that EW is either $<$ 3~m\AA\ or not computed in a model atmosphere with given parameters.
\end{table}

\begin{figure}
\centering
\includegraphics[trim={0 25 0 0},width=\hsize]{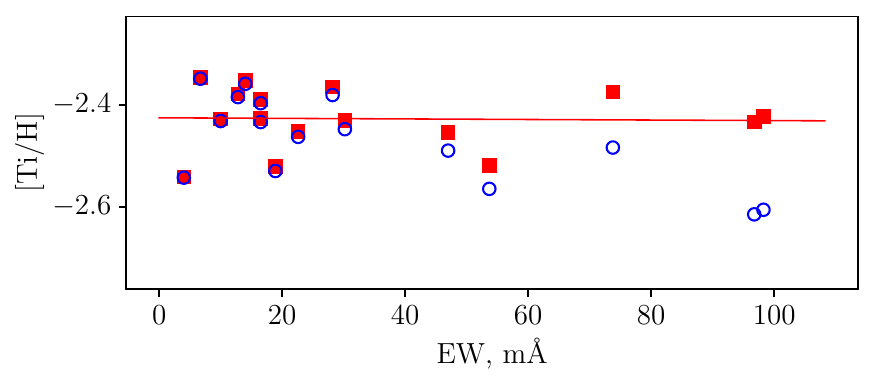}
\includegraphics[trim={0 10 0 0},width=\hsize]{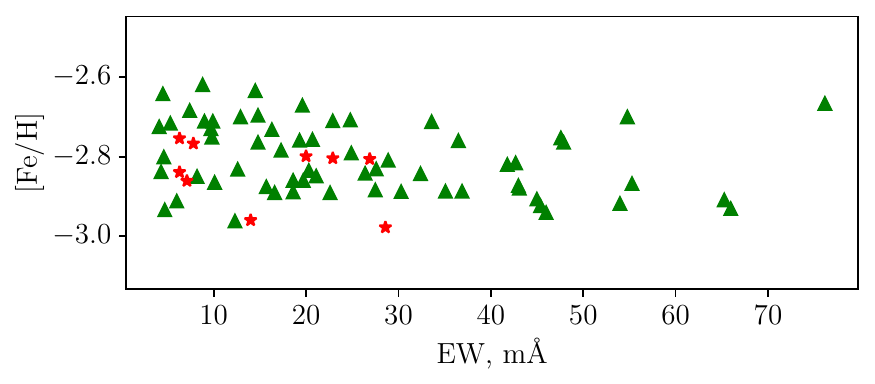}
\caption{Top panel: NLTE abundances from Ti\ii\ lines in CES1427--2214 as a function of the EW computed for \vt\ = 1.60~\kms\ (squares)  and  1.85~\kms\ (circles). Bottom panel: NLTE abundances from Fe\ione\ (triangles) and Fe\ii\ (asteriscs) lines in CES1427--2214 computed for \vt\ = 1.60~\kms.}
\label{vt_ew}
\end{figure}

When comparing the derived \vt\ values with those calculated from the empirical formula of MJ17, we found slightly lower values, with the difference never exceeding 0.25~\kms. For CES1427--2214, the star with the largest revision in \vt\ (0.25~\kms), we show the NLTE abundances from Ti\ii\ as a function of EW for \vt\ calculated using the MJ17 formula and those derived from Ti\ii\ lines (Fig.~\ref{vt_ew}). Using \vt\ = 1.85~\kms\ results in a significant trend in both NLTE and LTE, with slopes of $-2.0 \cdot 10^{-3}$ and $-1.3 \cdot 10^{-3}$ dex per m\AA, respectively. In contrast, a \vt\ value lower by 0.25~\kms\ results in consistent abundances from Ti\ii\ lines of different strengths, with slopes of $-5 \cdot 10^{-5}$ and $6 \cdot 10^{-4}$ dex per m\AA\ in NLTE and LTE, respectively.
For CES1427--2214, we ensure that the \vt\ derived from Ti\ii\ lines yields consistent NLTE abundances from Fe\ione\ and Fe\ii\ lines of different strength (Fig.~\ref{vt_ew}, bottom panel). See MJ17 for the line list and the NLTE method of iron abundance determination.

Observations show that stars with higher log~g have lower \vt\ \citep[see, for example, the empirical formula for \vt\ derived for dwarfs and subgiants in][]{lick}. The majority of our sample stars are giants at the base of the RGB, with log~g $\simeq$ 2.5, whereas the stellar sample of MJ17 consists mostly of cool giants, with 45 out of 59 stars having log~g less than 2.0. The difference in stellar parameters between our sample and the stars used to derive the empirical formula may explain the slightly lower \vt\ values obtained in this study. We find systematically lower \vt\ values in stars with higher log~g, whereas consistent \vt\ values are obtained for the four stars with log~g $<$ 2.0.

Using our new measurements, we derived an empirical formula suitable for \vt\ determination in giants at the base of the RGB:
\vt\ = --1.12 + 8.37$\cdot 10^{-4} \cdot$ \teff\ -- 6.48$\cdot 10^{-1} \cdot$ log~g.\\
Here, we assume that \vt\ does not depend on [Fe/H], as the corresponding coefficients in the MJ17 and \citet{lick} formulae are small and have opposite signs, suggesting no significant correlation between \vt\ and [Fe/H]. When comparing \vt\ measurements for individual stars with predictions from the above formula, we find that the difference does not exceed 0.10~\kms\ in absolute value, with a standard deviation of 0.05~\kms.

\subsection{Analysis of the Ba\ii\ lines}
We analyse five lines of Ba\ii: the strong resonance lines at $\lambda$ = 4554~\AA\ and 4934~\AA, and the subordinate lines at $\lambda$ = 5853~\AA, 6141~\AA, and 6496~\AA.

\begin{figure}
\centering
\includegraphics[trim={0 10 0 10},width=\hsize]{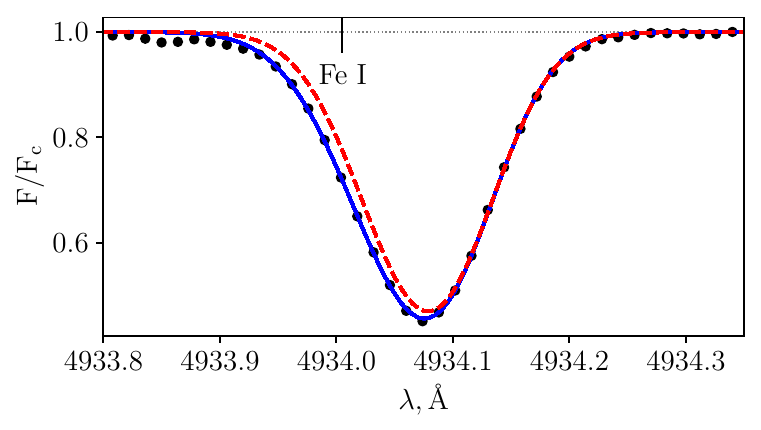}
\caption{Ba\ii\ 4934 \AA\ line profile in the observed spectrum (circles) of the most metal-rich sample star CES1436--2906 with [Fe/H] = $-2.15$ along with synthetic best-fit NLTE spectrum (solid line) and the spectrum calculated neglecting the Fe\ione\ 4934 \AA\ line (dashed line).}
\label{bafe4934}
\end{figure}

There is a Fe\ione\ 4934.0~\AA\ line with a lower level energy of \eexc\ = 4.15~eV and an oscillator strength of log~gf = $-0.58$ in the blue wing of the Ba\ii\ 4934.1~\AA\ line. Fifteen of our sample stars have [Fe/H] $\leq -2.4$, and blending with the Fe\ione\ 4934.0~\AA\ line is negligible. For the most metal-rich sample star, HD~128279 (CES1436--2906), with [Fe/H] = $-2.15$, we show the observed spectrum around the Ba\ii\ 4934.1~\AA\ line, along with synthetic spectra calculated with and without the iron line (Fig.~\ref{bafe4934}).

The subordinate line Ba\ii\ 6141.71~\AA\ is blended with the Fe\ione\ 6141.73~\AA\ line, which has \eexc\ = 3.60~eV and log~gf = $-1.46$. We account for this line in our spectrum synthesis using the iron abundance derived in LB22. For the most metal-rich star, varying the iron abundance within an uncertainty of 0.13~dex (LB22) results in negligible changes in the synthetic spectrum and does not affect barium abundance determination.

\subsubsection{NLTE effects}
To account for departures from LTE, we adopt the Ba\ii\ model atom from \citet{2019AstL...45..341M}. The NLTE effects for Ba\ii\ are moderate, leading to either weaker or stronger lines compared to LTE, depending on the line strength and stellar atmosphere parameters. In our sample stars, NLTE abundance corrections for individual Ba\ii\ lines do not exceed 0.13~dex in absolute value.

We draw attention to an intriguing line formation phenomenon observed for the Ba\ii\ resonance line in some stars. Based on the ratio of the oscillator strengths of the Ba\ii\ 4934~\AA\ and Ba\ii\ 4554~\AA\ lines (f$_{\rm 4934}$/f$_{\rm 4554}$ = 0.60), one might expect a similar ratio for their corresponding equivalent widths, assuming the spectral lines lie within the linear portion of the curve of growth. Although our lines are stronger and extend beyond the linear portion, in the three stars with the weakest Ba\ii\ lines, the ratio EW$_{\rm 4934}$/EW$_{\rm 4554}$ is approximately 0.7. Notably, this ratio increases for stronger Ba\ii\ lines, though it never reaches unity in LTE. In seven of our sample stars with strong resonance lines (EW $>$ 110~m\AA), the Ba\ii\ 4554~\AA\ and Ba\ii\ 4934~\AA\ lines have nearly the same strength, and in four of them, the EW of the Ba\ii\ 4934~\AA\ line slightly exceeds that of the Ba\ii\ 4554~\AA\ line.

\begin{figure}
\centering
\includegraphics[trim={0 10 0 10},width=\hsize]{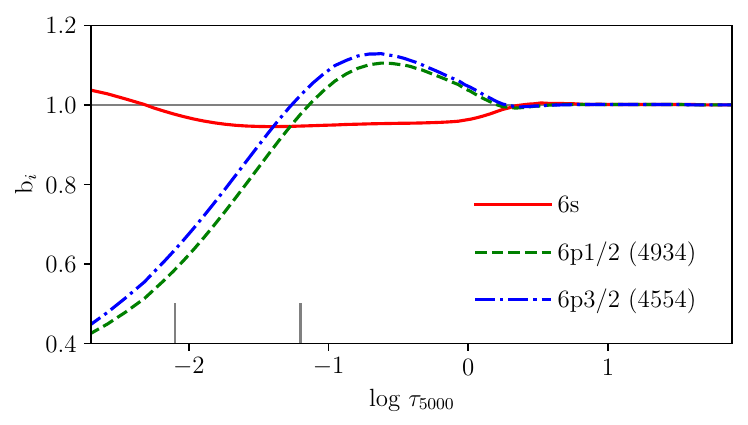}
\caption{Departure coefficients of the ground state of Ba\ii\ (6s) and the upper levels of the resolance transitions in a model atmosphere of a VMP giant. The vertical dashes indicate the line core formation depths of the resonance lines of different strength: 130~m\AA\ and 60~m\AA\ at log$\tau_{\rm 5000}$ = $-2.1$ and $-1.2$, respectively.}
\label{biba}
\end{figure}

When Ba\ii\ resonance lines are stronger than $\simeq$ 100~m\AA, NLTE leads to stronger lines compared to LTE, and the NLTE abundance corrections are negative. For the Ba\ii\ 4934~\AA\ line, NLTE corrections are larger in absolute value than for the Ba\ii\ 4554~\AA\ line. This paradox is explained by the fact that the NLTE effects for the Ba\ii\ 4934~\AA\ line result in more efficient line strengthening compared to the Ba\ii\ 4554~\AA\ line. Figure~\ref{biba} shows the departure coefficients for the ground state of Ba\ii\ (6s) and the upper levels of the resonance transitions in a model atmosphere of a VMP giant. The Ba\ii\ 4554~\AA\ and 4934~\AA\ lines form in the 6s~$-$~6p~(J~=~3/2) and 6s~$-$~6p~(J~=~1/2) transitions, respectively. 

Strong resonance lines form across a broad atmospheric depth range, up to log$\tau_{\rm 5000}$ = $-2.1$. In the high atmospheric layers, the upper levels are underpopulated relative to LTE and to the ground state, and the ratio S$_{\nu}$/B$_{\nu}$ $\sim$ b$_{\rm up}$/b$_{\rm low}$ $<$ 1, where S$_{\nu}$ and B$_{\nu}$ are the line source function and the Planck function, respectively. A decrease in the line source function results in a lower flux and a stronger line core compared to LTE. In deeper atmospheric layers with log$\tau_{\rm 5000}$ $> -1$, the situation is reversed: radiative pumping in the resonance transition results in a relative overpopulation of the upper levels relative to the ground state, and weakening the line compared to LTE. This effect is smaller for the Ba\ii\ 4934~\AA\ line, which has a lower oscillator strength than the Ba\ii\ 4554~\AA\ line. Due to the reduced line weakening in the deep layers, the Ba\ii\ 4934~\AA\ line may ultimately appear stronger than the Ba\ii\ 4554~\AA\ line.

\subsubsection{Impact of Ba isotope ratio on abundance determination}\label{subs_ba_iso_rat}
Barium abundance in stars can be represented by up to five isotopes: $^{\rm 134}$Ba, $^{\rm 135}$Ba, $^{\rm 136}$Ba, $^{\rm 137}$Ba, and the most abundant isotope, $^{\rm 138}$Ba. The isotopes $^{\rm 134}$Ba and $^{\rm 136}$Ba are s-only isotopes and cannot be produced in the r-process due to the presence of the stable r-only isotopes $^{\rm 134}$Xe and $^{\rm 136}$Xe, which block the $\beta$-decay path and prevent the formation of $^{\rm 134}$Ba and $^{\rm 136}$Ba \citep{1987RPPh...50..233R}.

Odd isotopes are subject to hyperfine splitting (HFS) of the energy levels, and a higher fraction of odd isotopes results in a broader line profile and greater total absorbed energy. HFS primarily affects the ground state, meaning that Ba\ii\ resonance lines can serve as a diagnostic of the Ba isotope ratio. In contrast, subordinate lines are unaffected by the adopted isotope ratio and can be used as reliable indicators of barium abundance. By determining the abundance from subordinate lines, one can infer the Ba isotope ratio, which can then be translated into the relative contributions of the r- and s-processes. The idea of estimating the r/s mixture of a star was first proposed by \citet{1989ApJ...346.1030C}, who pointed out the importance of accounting for HFS in Ba abundance determinations. This method has been applied to MP stars in the MW \citep{2006A&A...456..313M,2008A&A...478..529M,2019AstL...45..341M} and the Sculptor dwarf spheroidal (dSph) galaxy  \citep[][]{2015A&A...583A..67J}.

\citet{1993A&A...268L..27M} and \citet{1995A&A...297..686M} suggested a different method based on measuring the broadening of the Ba\ii\ resonance lines and applied it to the VMP star HD~140283. Depending on a quality of the observed spectrum and the adopted line formation calculation method, different autors report different results for Ba isotope origin in HD~140283: a pure r-process in 1D~LTE \citep{1993A&A...268L..27M,1995A&A...297..686M}, a solar mixture in 1D~LTE \citep[][]{2002MNRAS.335..325L} and in 3D~LTE \citep[][]{2009PASA...26..330C}, a pure s-process in 1D~LTE \citep[][]{2010A&A...523A..24G}, and a mostly r-process in 3D~LTE \citep{2015A&A...579A..94G}.

For barium, we determine abundances using different isotope ratios corresponding to the following r- to s-process fractions: 100\%/0\% (pure r-process), 80\%/20\%, 50\%/50\%, 12\%/88\% (solar), and 0\%/100\% (pure s-process). We adopt s-process and r-process residual Ba isotope yields from \citet[][]{2020MNRAS.491.1832P}. A larger r-process fraction results in a higher proportion of odd isotopes and a lower abundance derived from the resonance lines compared to that obtained with a lower r-process fraction.
For each star, we provide abundances from individual spectral lines along with their measured EWs (Table~\ref{ba_lbl}).

\begin{table*}
\caption{NLTE and LTE abundances and EWs (m\AA) of Ba\ii\ lines in the sample stars.}
\label{ba_lbl}
\setlength{\tabcolsep}{0.90mm}
\centering   
   \begin{tabular}{lrrrrrrrrrrrrrr}    
      \hline
 & Ba\ii\ lines:    & \multicolumn{5}{c}{  4554 \AA} & \multicolumn{5}{c}{  4934 \AA} &  5853 \AA &  6141 \AA & 6496 \AA \\
Star &   & \multicolumn{5}{c}{---------------------------------------} & \multicolumn{5}{c}{---------------------------------------} &    &    &   \\
& r/s ratio: & 100/0 & 80/20 & 50/50 & 15/85 & 0/100 & 100/0 & 80/20 & 50/50 & 12/88 & 0/100 & & &  \\
   \hline 
       CES1322--1355 & NLTE &   --1.06 &   --1.02 &   --0.94 &   --0.76 &   --0.64 &   --1.11 &   --1.07 &   --1.00 &   --0.83 &   --0.72 &   --0.95 &   --0.96 &   --0.95 \\ 
       CES1322--1355 & LTE  &   --1.07 &   --1.03 &   --0.95 &   --0.77 &   --0.65 &   --1.13 &   --1.09 &   --1.02 &   --0.85 &   --0.74 &   --1.02 &   --1.04 &   --0.99 \\ 
       CES1322--1355 & EW   &   111.0 &   111.0 &   111.0 &   111.0 &   111.0 &   102.1 &   102.1 &   102.1 &   102.1 &   102.1 &    11.2 &    47.0 &    40.0 \\ 
       CES1436--2906 & NLTE &   --0.60 &   --0.56 &   --0.48 &   --0.31 &   --0.19 &   --0.65 &   --0.62 &   --0.56 &   --0.42 &   --0.33 &   --0.39 &   --0.48 &   --0.36 \\ 
       CES1436--2906 & LTE  &   --0.58 &   --0.54 &   --0.46 &   --0.29 &   --0.17 &   --0.66 &   --0.63 &   --0.57 &   --0.43 &   --0.34 &   --0.45 &   --0.54 &   --0.37 \\ 
       CES1436--2906 & EW   &    85.8 &    85.8 &    85.8 &    85.8 &    85.8 &    74.0 &    74.0 &    74.0 &    74.0 &    74.0 &     8.2 &    34.0 &    31.5 \\ 
       CES1427--2214 & NLTE &   --0.78 &   --0.74 &   --0.66 &   --0.48 &   --0.35 &   --0.84 &   --0.80 &   --0.71 &   --0.51 &   --0.37 &   --0.77 &   --0.77 &   --0.73 \\ 
       CES1427--2214 & LTE  &   --0.75 &   --0.71 &   --0.63 &   --0.45 &   --0.32 &   --0.81 &   --0.77 &   --0.68 &   --0.48 &   --0.34 &   --0.83 &   --0.83 &   --0.76 \\ 
       CES1427--2214 & EW   &   115.8 &   115.8 &   115.8 &   115.8 &   115.8 &   113.4 &   113.4 &   113.4 &   113.4 &   113.4 &    15.5 &    53.8 &    48.5 \\ 
        CES1402+0941 & NLTE &   --1.66 &   --1.63 &   --1.56 &   --1.41 &   --1.31 &   --1.67 &   --1.65 &   --1.60 &   --1.48 &   --1.40 &  -- &   --1.41 &   --1.40 \\ 
        CES1402+0941 & LTE  &   --1.70 &   --1.67 &   --1.60 &   --1.45 &   --1.35 &   --1.69 &   --1.67 &   --1.62 &   --1.50 &   --1.42 &  -- &   --1.50 &   --1.43 \\ 
        CES1402+0941 & EW   &    99.6 &    99.6 &    99.6 &    99.6 &    99.6 &    90.4 &    90.4 &    90.4 &    90.4 &    90.4 &   -- &    41.0 &    34.9 \\ 
        CES1226+0518 & NLTE &   --0.13 &   --0.10 &   --0.03 &    0.14 &    0.25 &   --0.19 &   --0.15 &   --0.07 &    0.11 &    0.24 &   --0.05 &  -- &   --0.08 \\ 
        CES1226+0518 & LTE  &   --0.06 &   --0.03 &    0.04 &    0.21 &    0.32 &   --0.10 &   --0.06 &    0.02 &    0.20 &    0.33 &   --0.09 &  -- &   --0.04 \\ 
        CES1226+0518 & EW   &   114.1 &   114.1 &   114.1 &   114.1 &   114.1 &   112.9 &   112.9 &   112.9 &   112.9 &   112.9 &    19.5 &   -- &    52.6 \\ 
        CES1732+2344 & NLTE &   --0.87 &   --0.84 &   --0.77 &   --0.61 &   --0.50 &   --0.87 &   --0.84 &   --0.79 &   --0.67 &   --0.59 &  -- &  -- &   --0.70 \\ 
        CES1732+2344 & LTE  &   --0.91 &   --0.88 &   --0.81 &   --0.65 &   --0.54 &   --0.90 &   --0.87 &   --0.82 &   --0.70 &   --0.62 &  -- &  -- &   --0.75 \\ 
        CES1732+2344 & EW   &    75.6 &    75.6 &    75.6 &    75.6 &    75.6 &    65.4 &    65.4 &    65.4 &    65.4 &    65.4 &   -- &   -- &    20.1 \\ 
        CES0301+0616 & NLTE &   --0.11 &   --0.08 &   --0.01 &    0.13 &    0.23 &   --0.12 &   --0.08 &    0.00 &    0.18 &    0.29 &   --0.15 &   --0.17 &   --0.17 \\ 
        CES0301+0616 & LTE  &   --0.03 &    0.00 &    0.07 &    0.21 &    0.31 &   --0.02 &    0.02 &    0.10 &    0.28 &    0.39 &   --0.18 &   --0.15 &   --0.14 \\ 
        CES0301+0616 & EW   &   113.1 &   113.1 &   113.1 &   113.1 &   113.1 &   114.0 &   114.0 &   114.0 &   114.0 &   114.0 &    18.2 &    53.5 &    47.8 \\ 
       CES0338--2402 & NLTE &   --1.24 &   --1.21 &   --1.16 &   --1.03 &   --0.94 &   --1.23 &   --1.21 &   --1.18 &   --1.09 &   --1.03 &   --1.16 &   --1.04 &   --1.05 \\ 
       CES0338--2402 & LTE  &   --1.31 &   --1.28 &   --1.23 &   --1.10 &   --1.01 &   --1.29 &   --1.27 &   --1.24 &   --1.15 &   --1.09 &   --1.23 &   --1.14 &   --1.09 \\ 
       CES0338--2402 & EW   &    59.1 &    59.1 &    59.1 &    59.1 &    59.1 &    45.3 &    45.3 &    45.3 &    45.3 &    45.3 &     2.6 &    17.0 &    12.9 \\ 
       CES0422--3715 & NLTE &   --0.30 &   --0.26 &   --0.19 &   --0.03 &    0.07 &   --0.30 &   --0.26 &   --0.18 &    0.00 &    0.11 &   --0.21 &  -- &   --0.28 \\ 
       CES0422--3715 & LTE  &   --0.23 &   --0.19 &   --0.12 &    0.04 &    0.14 &   --0.20 &   --0.16 &   --0.08 &    0.10 &    0.21 &   --0.24 &  -- &   --0.23 \\ 
       CES0422--3715 & EW   &   123.8 &   123.8 &   123.8 &   123.8 &   123.8 &   125.2 &   125.2 &   125.2 &   125.2 &   125.2 &    25.4 &   -- &    60.0 \\ 
       CES2231--3238 & NLTE &   --0.04 &    0.00 &    0.06 &    0.22 &    0.32 &   --0.08 &   --0.04 &    0.04 &    0.23 &    0.36 &   --0.06 &   --0.16 &   --0.15 \\ 
       CES2231--3238 & LTE  &    0.06 &    0.10 &    0.16 &    0.32 &    0.42 &    0.05 &    0.09 &    0.17 &    0.36 &    0.49 &   --0.09 &   --0.12 &   --0.09 \\ 
       CES2231--3238 & EW   &   125.5 &   125.5 &   125.5 &   125.5 &   125.5 &   126.3 &   126.3 &   126.3 &   126.3 &   126.3 &    25.4 &    62.6 &    57.9 \\ 
       CES2232--4138 & NLTE &   --0.61 &   --0.58 &   --0.51 &   --0.34 &   --0.23 &   --0.65 &   --0.62 &   --0.55 &   --0.38 &   --0.28 &   --0.48 &  -- &   --0.47 \\ 
       CES2232--4138 & LTE  &   --0.55 &   --0.52 &   --0.45 &   --0.28 &   --0.17 &   --0.60 &   --0.57 &   --0.50 &   --0.33 &   --0.23 &   --0.53 &  -- &   --0.48 \\ 
       CES2232--4138 & EW   &    99.3 &    99.3 &    99.3 &    99.3 &    99.3 &    92.7 &    92.7 &    92.7 &    92.7 &    92.7 &    11.0 &   -- &    38.4 \\ 
       CES0419--3651 & NLTE &   --1.46 &   --1.43 &   --1.37 &   --1.24 &   --1.15 &   --1.49 &   --1.47 &   --1.43 &   --1.35 &   --1.29 &  -- &   --1.33 &   --1.22 \\ 
       CES0419--3651 & LTE  &   --1.53 &   --1.50 &   --1.44 &   --1.31 &   --1.22 &   --1.56 &   --1.54 &   --1.50 &   --1.42 &   --1.36 &  -- &   --1.42 &   --1.26 \\ 
       CES0419--3651 & EW   &    64.1 &    64.1 &    64.1 &    64.1 &    64.1 &    46.1 &    46.1 &    46.1 &    46.1 &    46.1 &   -- &    17.4 &    15.7 \\ 
       CES0045--0932 & NLTE &   --1.60 &   --1.58 &   --1.53 &   --1.41 &   --1.34 &   --1.58 &   --1.57 &   --1.54 &   --1.47 &   --1.43 &   --1.41 &   --1.39 &   --1.37 \\ 
       CES0045--0932 & LTE  &   --1.67 &   --1.65 &   --1.60 &   --1.48 &   --1.41 &   --1.64 &   --1.63 &   --1.60 &   --1.53 &   --1.49 &   --1.48 &   --1.47 &   --1.41 \\ 
       CES0045--0932 & EW   &    59.2 &    59.2 &    59.2 &    59.2 &    59.2 &    42.2 &    42.2 &    42.2 &    42.2 &    42.2 &     2.6 &    15.8 &    12.9 \\ 
       CES0215--2554 & NLTE &   --1.08 &   --1.04 &   --0.98 &   --0.82 &   --0.72 &   --1.09 &   --1.06 &   --1.01 &   --0.87 &   --0.79 &   --0.89 &   --0.89 &   --0.90 \\ 
       CES0215--2554 & LTE  &   --1.09 &   --1.05 &   --0.99 &   --0.83 &   --0.73 &   --1.10 &   --1.07 &   --1.02 &   --0.88 &   --0.80 &   --0.96 &   --0.97 &   --0.94 \\ 
       CES0215--2554 & EW   &   100.9 &   100.9 &   100.9 &   100.9 &   100.9 &    84.3 &    84.3 &    84.3 &    84.3 &    84.3 &     9.2 &    41.6 &    33.7 \\ 
       CES2330--5626 & NLTE &   --0.17 &   --0.14 &   --0.07 &    0.07 &    0.17 &   --0.20 &   --0.16 &   --0.08 &    0.10 &    0.23 &   --0.31 &   --0.32 &   --0.32 \\ 
       CES2330--5626 & LTE  &   --0.07 &   --0.04 &    0.03 &    0.17 &    0.27 &   --0.09 &   --0.05 &    0.03 &    0.21 &    0.34 &   --0.34 &   --0.27 &   --0.26 \\ 
       CES2330--5626 & EW   &   133.2 &   133.2 &   133.2 &   133.2 &   133.2 &   134.9 &   134.9 &   134.9 &   134.9 &   134.9 &    27.0 &    69.2 &    63.9 \\ 
        CES1222+1136 & NLTE &   --0.69 &   --0.65 &   --0.57 &   --0.40 &   --0.28 &   --0.78 &   --0.74 &   --0.66 &   --0.47 &   --0.34 &   --0.72 &   --0.73 &   --0.70 \\ 
        CES1222+1136 & LTE  &   --0.63 &   --0.59 &   --0.51 &   --0.34 &   --0.22 &   --0.71 &   --0.67 &   --0.59 &   --0.40 &   --0.27 &   --0.77 &   --0.77 &   --0.70 \\ 
        CES1222+1136 & EW   &   133.5 &   133.5 &   133.5 &   133.5 &   133.5 &   129.7 &   129.7 &   129.7 &   129.7 &   129.7 &    22.9 &    66.3 &    62.5 \\ 
\hline
      \end{tabular}\\
\end{table*}

\begin{figure}
\centering
\includegraphics[trim={0 40 0 0},width=\hsize]{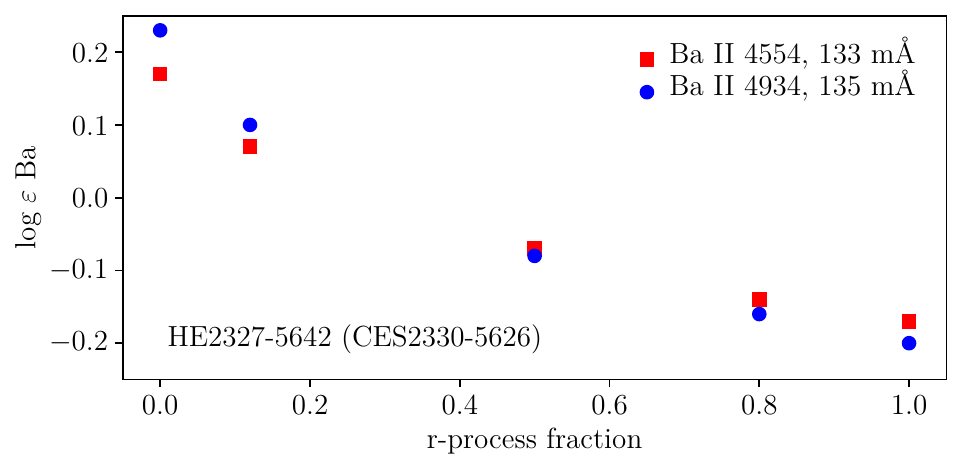}
\includegraphics[trim={0 10 0 0},width=\hsize]{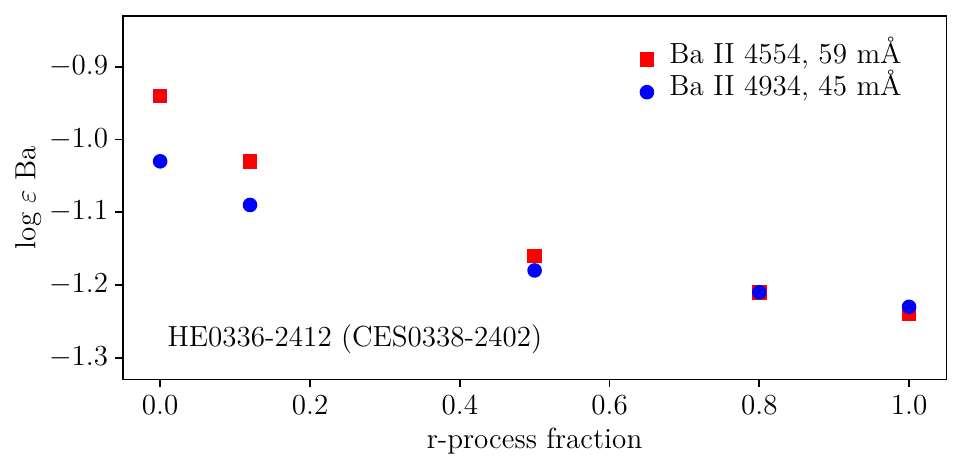}
\caption{The NLTE abundances from the Ba\ii\ 4554~\AA\ and 4934 \AA\ lines as a function of r-process contribution to Ba isotopes in the two sample stars. The EWs of the resonance lines are indicated.}
\label{ba_res_rfr}
\end{figure}

Figure~\ref{ba_res_rfr} illustrates the impact of the adopted Ba isotope mixture on barium abundance derived from the resonance lines. We plot the abundances from Ba\ii\ 4554~\AA\ and 4934~\AA\ lines in two sample stars: one with the weakest lines (HE0336--2412/CES0338--2402) and another with the strongest lines (r-\ii\ type star HE2327--5642/CES2330--5626). The stronger the Ba\ii\ resonance lines, the more sensitive their abundances are to the adopted Ba isotope mixture. For example, for the Ba\ii\ 4554~\AA\ line with EW = 59~m\AA, the abundance difference between pure r- and pure s-process Ba isotope mixtures is 0.30~dex. The corresponding difference in HE2327--5642, where EW$_{\rm 4554}$ = 133~m\AA, amounts to 0.34~dex. Notably, in HE2327--5642, the Ba\ii\ 4934~\AA\ line is slightly more sensitive to the Ba isotope mixture, as it is slightly stronger than the Ba\ii\ 4554~\AA\ line due to the effect discussed earlier.

\begin{figure}
\centering
\includegraphics[trim={0 40 0 0},width=\hsize]{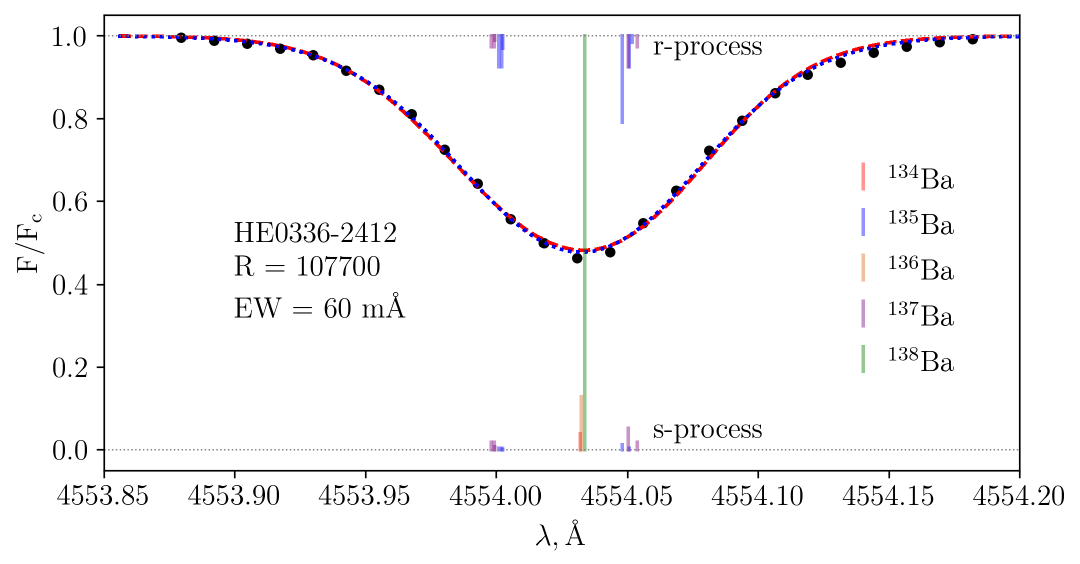}
\includegraphics[trim={0 0 0 0},width=\hsize]{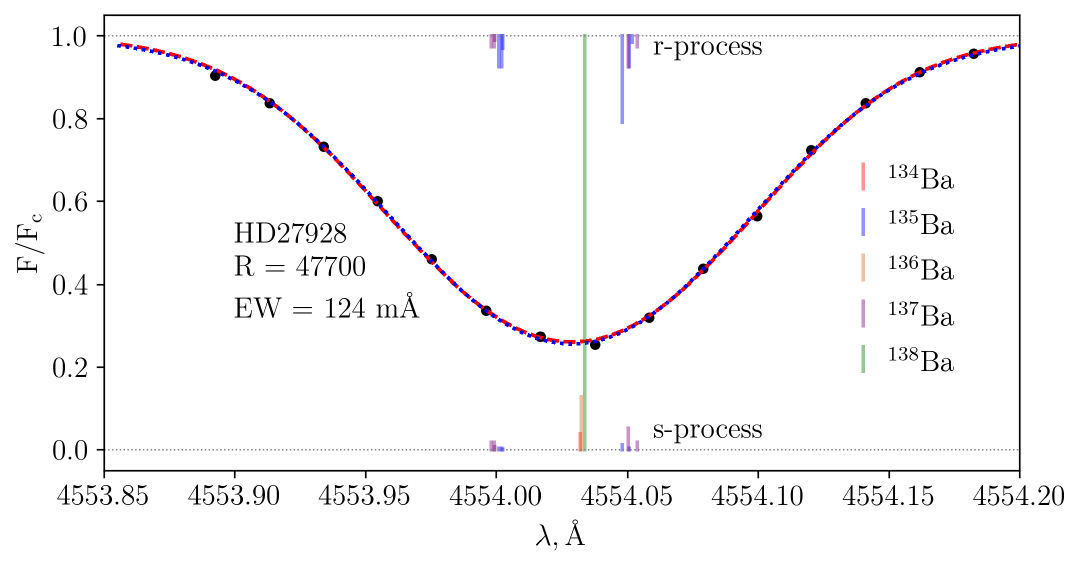}
\caption{Top panel: Ba\ii\ 4554 \AA\ line profile in the observed spectrum (circles) of HE0336--2412 (CES0338--2402). Synthetic best-fit NLTE spectra, derived using pure r-process (dashed line) and pure s-process (dotted line) Ba isotope mixtures are almost indistinguishable, demonstrating that a reasonable fit can be achieved with any adopted isotope ratio. Vertical dashes show the relative contribution of differend Ba isotopes to the r-prosess and s-process, see legend for designations. Bottom panel: The same as the top panel for HD~27928 (CES0422--3715).}
\label{fig_ba4554_rs}
\end{figure}

The adopted Ba isotope mixture has a significant impact on the abundance derived from resonance lines, while the observed line profiles can be fitted when adopting any Ba isotope mixture. To demonstrate the quality of spectral fitting, we selected the UVES spectrum of HE0336--2412 (CES0338--2402), which has a spectral resolution of R = 107200, higher than that of most sample stars, whose spectra were observed at R $\simeq$ 45000. Since HE0336--2412 has one of the weakest Ba\ii\ 4554~\AA\ lines (EW = 59~m\AA), we also include the spectrum of HD~27928 (CES0422--3715) with EW$_{\rm 4554}$ = 124~m\AA, obtained using HIRES with R = 47700 (Fig.~\ref{fig_ba4554_rs}). For illustration, we present the best-fit synthetic NLTE spectra of the Ba\ii\ 4554~\AA\ line computed with pure r-process and pure s-process Ba isotope ratios. Regardless of whether an r- or s-process Ba isotope mixture is used, we achieve a satisfactory fit to the observed spectra.

For CES0338--2402, the best-fit spectrum parameters are $\eps$(Ba)/v$_{\rm mac}$ = $-1.24$/3.6~\kms\ (pure r-process) and $-0.94$/4.2~\kms\ (pure s-process). The corresponding values for CES0422--3715 are $\eps$(Ba)/v$_{\rm mac}$ = $-0.30$/3.0~\kms\ and $0.07$/3.3~\kms. In our abundance determination method, v$_{\rm mac}$ is treated as a free parameter when fitting individual spectral lines. From analysis of different spectral lines in a given star, we estimate a typical uncertainty in v$_{\rm mac}$ as 2~\kms. This uncertainty prevents us from determining the Ba isotope mixture solely from the Ba\ii\ resonance line profiles. 

\subsection{Sr and Eu abundances}
We determine Sr abundances from the Sr\ii\ 4077~\AA\ and 4215~\AA\ lines. To account for the NLTE effects, we rely on the Sr\ii\ model atom presented in \citet{sr_nlte} and the NLTE abundance corrections from the INASAN database\footnote{https://spectrum.inasan.ru/nLTE2/} \citep{2023MNRAS.524.3526M}. There are two reasons why we determine Sr abundances in this study instead of taking them from LB22. First, for six of the selected sample stars, Sr abundances are missing in LB22. 
Secondly, a revision in \vt\ affects the Sr\ii\ lines, as they are strong in our sample stars, and Sr abundances might be up to 0.2~dex higher than those computed with the \vt\ from the MJ17 formula.

We determine Eu abundances in our sample stars using the Eu\ii\ model atom of \citet{2000AA...364..249M}. LH25 relies on the Eu\ii\ 3819~\AA\ and 4129~\AA\ lines. The first of them is difficult for analysis, since this line is located in the blue wing of a strong Fe\ione\ 3819.62~\AA\ line and, in addition to that, its blue wing is blended with the Cr\ione\ 3819.56~\AA\ line. In stars with weak Eu lines, the above uncertainty, along with the uncertainty in the continuum placement, results in a systematically lower abundance from the Eu\ii\ 3819~\AA\ compared to that from the Eu 4129~\AA\ line. Instead of this line, we included in our analysis the Eu\ii\ 4205~\AA\ line, which is free of blending in the metallicity range we deal with in this study. The average Eu abundances are calculated from Eu\ii\ 4129~\AA\ and 4205~\AA\ lines. Three of the sample stars have very weak lines of Eu\ii\ with EW $<$ 6~m\AA, and we use their Eu abundances with caution. These measurements are marked with a $<$ symbol in Table~\ref{ratios_table}.

\begin{table}
\caption{NLTE and LTE abundances and EWs (m\AA) from individual Sr\ii\ and Eu\ii\ lines in the sample stars.}
\label{sr_eu_lbl}
\setlength{\tabcolsep}{0.90mm}
\centering   
   \begin{tabular}{lrrrrrr}    
      \hline
Star  &  & \multicolumn{2}{c}{Sr\ii} & \multicolumn{3}{c}{Eu\ii} \\
  &  & \multicolumn{2}{c}{----------------} & \multicolumn{3}{c}{------------------------} \\
 & $\lambda$, \AA :   &  4077 &  4215 &  3819 &    4129 &   4205 \\
\hline
      CES1322--1355  & NLTE &    0.32 &    0.21 &   --2.07 &   --1.95 &   --1.94 \\ 
      CES1322--1355  & LTE  &    0.35 &    0.26 &   --2.25 &   --2.10 &   --2.08 \\ 
      CES1322--1355  & EW   &   167.9 &   144.4 &    19.6 &    15.1 &    16.3 \\ 
      CES1436--2906  & NLTE &    0.14 &    0.14 &   --1.58 &   --1.55 &   --1.47 \\ 
      CES1436--2906  & LTE  &    0.22 &    0.23 &   --1.69 &   --1.65 &   --1.57 \\ 
      CES1436--2906  & EW   &   108.4 &    96.8 &    14.4 &     8.4 &    11.7 \\ 
      CES1427--2214  & NLTE &   --0.01 &   --0.06 &   --1.84 &   --1.80 &   --1.76 \\ 
      CES1427--2214  & LTE  &    0.03 &    0.00 &   --1.99 &   --1.94 &   --1.89 \\ 
      CES1427--2214  & EW   &   136.4 &   121.1 &    31.6 &    20.4 &    21.4 \\ 
       CES1402+0941  & NLTE &    0.10 &   --0.01 &   --2.74 &   --2.52 &   --2.50 \\ 
       CES1402+0941  & LTE  &    0.08 &   --0.01 &   --2.96 &   --2.71 &   --2.66 \\ 
       CES1402+0941  & EW   &   178.8 &   154.6 &    10.9 &     8.8 &    14.1 \\ 
       CES1226+0518  & NLTE &    0.51 &    0.42 &   --1.13 &   --1.07 &   --1.05 \\ 
       CES1226+0518  & LTE  &    0.59 &    0.54 &   --1.24 &   --1.18 &   --1.15 \\ 
       CES1226+0518  & EW   &   137.1 &   117.9 &    40.6 &    26.8 &    26.4 \\ 
       CES1732+2344  & NLTE &    0.14 &    0.14 &   --1.73 &   --1.55 &   --1.59 \\ 
       CES1732+2344  & LTE  &    0.24 &    0.26 &   --1.86 &   --1.67 &   --1.70 \\ 
       CES1732+2344  & EW   &   107.9 &    97.9 &    12.6 &     9.4 &     9.3 \\ 
       CES0301+0616  & NLTE &    0.27 &    0.23 &   --1.14 &   --1.09 &   --1.03 \\ 
       CES0301+0616  & LTE  &    0.35 &    0.33 &   --1.23 &   --1.18 &   --1.12 \\ 
       CES0301+0616  & EW   &   125.2 &   106.9 &    48.8 &    33.1 &    29.0 \\ 
      CES0338--2402  & NLTE &   --0.10 &   --0.20 &   --2.39 &   --2.13 &   --2.15 \\ 
      CES0338--2402  & LTE  &   --0.02 &   --0.11 &   --2.53 &   --2.25 &   --2.27 \\ 
      CES0338--2402  & EW   &   107.3 &    93.7 &     3.9 &     3.3 &     4.2 \\ 
      CES0422--3715  & NLTE &    0.31 &    0.27 &   --1.24 &   --1.17 &   --1.14 \\ 
      CES0422--3715  & LTE  &    0.36 &    0.35 &   --1.35 &   --1.28 &   --1.24 \\ 
      CES0422--3715  & EW   &   142.1 &   125.3 &    55.7 &    40.3 &    39.5 \\ 
      CES2231--3238  & NLTE &    0.30 &    0.17 &   --0.99 &   --0.94 &   --0.94 \\ 
      CES2231--3238  & LTE  &    0.37 &    0.27 &   --1.08 &   --1.04 &   --1.03 \\ 
      CES2231--3238  & EW   &   131.2 &   111.3 &    67.6 &    48.1 &    47.5 \\ 
      CES2232--4138  & NLTE &    0.44 &    0.39 &   --1.49 &   --1.45 &   --1.37 \\ 
      CES2232--4138  & LTE  &    0.49 &    0.47 &   --1.61 &   --1.56 &   --1.47 \\ 
      CES2232--4138  & EW   &   141.7 &   122.4 &    25.5 &    16.2 &    20.0 \\ 
      CES0419--3651  & NLTE &   --0.08 &   --0.09 &   --2.80 &   --2.57 &   --2.58 \\ 
      CES0419--3651  & LTE  &   --0.02 &   --0.02 &   --2.96 &   --2.71 &   --2.71 \\ 
      CES0419--3651  & EW   &   121.1 &   110.1 &     3.0 &     1.1 &     5.5 \\ 
      CES0045--0932  & NLTE &   --0.16 &   --0.17 &   --2.64 &   --2.35 &   --2.29 \\ 
      CES0045--0932  & LTE  &   --0.10 &   --0.10 &   --2.80 &   --2.49 &   --2.42 \\ 
      CES0045--0932  & EW   &   120.2 &   109.4 &     4.1 &     4.0 &     6.1 \\ 
      CES0215--2554  & NLTE &   --0.11 &   --0.19 &   --1.94 &   --1.87 &   --1.91 \\ 
      CES0215--2554  & LTE  &   --0.06 &   --0.13 &   --2.11 &   --2.01 &   --2.04 \\ 
      CES0215--2554  & EW   &   133.1 &   119.4 &    19.2 &    13.0 &    14.6 \\ 
      CES2330--5626  & NLTE &    0.11 &    0.09 &   --1.22 &   --1.18 &   --1.20 \\ 
      CES2330--5626  & LTE  &    0.16 &    0.16 &   --1.32 &   --1.28 &   --1.29 \\ 
      CES2330--5626  & EW   &   119.5 &   119.6 &    71.6 &    51.8 &    51.8 \\ 
       CES1222+1136  & NLTE &    0.11 &    0.02 &   --1.79 &   --1.71 &   --1.73 \\ 
       CES1222+1136  & LTE  &    0.13 &    0.06 &   --1.94 &   --1.85 &   --1.85 \\ 
       CES1222+1136  & EW   &   156.3 &   135.7 &    45.9 &    32.8 &    35.3 \\ 
 \hline
      \end{tabular}\\
The table is accessible in a machine-readable format at the CDS. A portion is shown to illustrate its format and content. 
\end{table}

\subsection{Element abundances and uncertainties}
Table~\ref{ratios_table} presents the NLTE abundance ratios and fractions of odd Ba isotopes F$_{\rm odd}$ = (N($^{135}$Ba)+N($^{137}$Ba))/N(Ba). We adopt solar abundances from \citet{2021SSRv..217...44L}: $\eps$(Sr)$_{\rm \odot}$ = 2.88, $\eps$(Ba)$_{\rm \odot}$ = 2.17, $\eps$(Eu)$_{\rm \odot}$ = 0.51.

Uncertainties in [X/H] ratios are computed, including uncertainties in stellar atmosphere parameters along with the dispersion of the single line measurements around the mean $\sigma_{\rm st} = \sqrt{ \Sigma (\eps - \eps_i )^2 /(N - 1)}$, where N is the total number of lines. For our Sr, Ba, and Eu abundance determination, we assume that the dispersion and the systematic uncertainties caused by variations in stellar parameters are uncorrelated. This assumption is valid since, for each of the studied elements, their spectral lines belong to the same species, have nearly the same intensity, and originate from atomic levels with similar or identical \eexc. Therefore, changes in \teff, log~g, and \vt\  do not contribute to line-to-line scatter, allowing the total abundance ratio [X/H]  uncertainty, to be calculated as follows:\\
$\sigma^2_{\rm [X/H]} = \sigma^2(X) + \sigma^2(T_{\rm eff}) + \sigma^2(\log~g) + \sigma^2(\xi_{\rm t})$

When computing abundance ratios [X/Y], where X and Y are Sr, Ba, and Eu, we neglect the uncertainty in log~g, since a shift in log~g produces the same abundance shift for the above elements. The corresponding uncertainties caused by changes in \teff\ and \vt\ are reduced compared to those for absolute abundances, since changes in these parameters affect the Sr\ii , Ba\ii , and Eu\ii\ spectral lines in the same way. For abundance ratios X and Y, the uncertainty in their ratio is calculated as follows:\\
$\sigma^2_{\rm [X/Y]} = \sigma^2(X) + \sigma^2(Y) + \sigma^2(\Delta T_{\rm eff}) + \sigma^2(\Delta \xi_{\rm t})$

\begin{table*}
\caption{NLTE abundance ratios and Ba odd isotope fractions of sample stars}
\setlength{\tabcolsep}{0.90mm}
 \label{ratios_table}
\centering   
   \begin{tabular}{cccrrrrrl}    
      \hline
Name & [Sr/H] &  [Ba/H]  & [Eu/H] &  [Sr/Ba] & [Ba/Eu] & [Sr/Eu] & [Eu/Fe] & F$_{\rm odd}$ \\ 
   \hline 
CES0419--3651 &  --2.97 (0.13) &  --3.45 (0.09) & $<$ --3.09 (0.08) &    0.48 (0.11) & $>$ --0.36 (0.06) & $>$~~~0.12 (0.09) & $<$ --0.30 & 0.15$^{+0.15}_{-0.05}$\\ 
CES1402+0941  &  --2.83 (0.14) &  --3.58 (0.07) &     --3.02 (0.08) &    0.74 (0.11) &     --0.55 (0.02) &       0.19 (0.11) &     --0.25 & 0.14$^{+0.09}_{-0.04}$\\ 
CES0045--0932 &  --3.05 (0.13) &  --3.56 (0.07) & $<$ --2.83 (0.08) &    0.51 (0.09) & $>$ --0.73 (0.04) & $>$ --0.21 (0.10) & $<$   0.10 & 0.11$^{+0.12}_{-0.01}$\\ 
CES1436--2906 &  --2.74 (0.13) &  --2.58 (0.08) &     --2.02 (0.08) &  --0.15 (0.11) &     --0.56 (0.06) &     --0.71 (0.10) &       0.11  & 0.23$^{+0.13}_{-0.08}$\\
CES0338--2402 &  --3.03 (0.14) &  --3.25 (0.09) & $<$ --2.65 (0.08) &    0.22 (0.12) & $>$ --0.60 (0.06) & $>$ --0.38 (0.10) & $<$   0.14 & 0.20$^{+0.18}_{-0.08}$\\
CES0215--2554 &  --3.03 (0.14) &  --3.06 (0.07) &     --2.40 (0.08) &    0.03 (0.10) &     --0.67 (0.02) &     --0.64 (0.10) &       0.31 & 0.23$^{+0.13}_{-0.07}$\\
CES1322--1355 &  --2.62 (0.14) &  --3.12 (0.07) &     --2.45 (0.08) &    0.51 (0.11) &     --0.67 (0.01) &     --0.16 (0.11) &       0.46 & 0.39$^{+0.20}_{-0.13}$\\ 
CES1732+2344  &  --2.74 (0.13) &  --2.87 (0.08) &     --2.08 (0.08) &    0.13 (0.10) &     --0.79 (0.05) &     --0.66 (0.09) &       0.47 & 0.26$^{+0.27}_{-0.13}$\\ 
CES2232--4138 &  --2.47 (0.13) &  --2.65 (0.07) &     --1.92 (0.09) &    0.18 (0.10) &     --0.73 (0.04) &     --0.55 (0.10) &       0.64 & 0.32$^{+0.16}_{-0.10}$\\ 
CES1222+1136  &  --2.81 (0.14) &  --2.89 (0.07) &     --2.23 (0.08) &    0.07 (0.10) &     --0.66 (0.02) &     --0.58 (0.10) &       0.66 & 0.69$^{+0.06}_{-0.23}$\\ 
CES1427--2214 &  --2.92 (0.13) &  --2.93 (0.07) &     --2.29 (0.08) &    0.01 (0.10) &     --0.63 (0.03) &     --0.62 (0.10) &       0.74 & 0.58$^{+0.17}_{-0.17}$\\ 
CES0422--3715 &  --2.59 (0.13) &  --2.42 (0.08) &     --1.67 (0.08) &  --0.18 (0.10) &     --0.75 (0.04) &     --0.92 (0.10) &       0.76 & 0.57$^{+0.18}_{-0.18}$\\ 
CES1226+0518  &  --2.41 (0.14) &  --2.24 (0.07) &     --1.57 (0.08) &  --0.18 (0.10) &     --0.67 (0.02) &     --0.84 (0.10) &       0.79 & 0.46$^{+0.24}_{-0.15}$\\ 
CES2330--5626 &  --2.78 (0.13) &  --2.49 (0.07) &     --1.70 (0.08) &  --0.30 (0.09) &     --0.79 (0.01) &     --1.08 (0.09) &       1.38 & 0.75\\ 
CES0301+0616  &  --2.63 (0.13) &  --2.33 (0.07) &     --1.57 (0.08) &  --0.30 (0.10) &     --0.77 (0.03) &     --1.06 (0.10) &       1.34 & 0.75$_{-0.11}$\\ 
CES2231--3238 &  --2.65 (0.15) &  --2.29 (0.08) &     --1.45 (0.08) &  --0.35 (0.12) &     --0.84 (0.05) &     --1.19 (0.11)  &       1.30 & 0.75$_{-0.06}$\\ 
 \hline
      \end{tabular}\\
Solar abundances are taken from \citet{2021SSRv..217...44L}. Total uncertainties are given in parenthesis. 
\end{table*}

\section{Discussion}\label{discussion}
\subsection{Abundance trends}
\begin{figure}
\centering
\includegraphics[trim={0 10 0 5},width=\hsize]{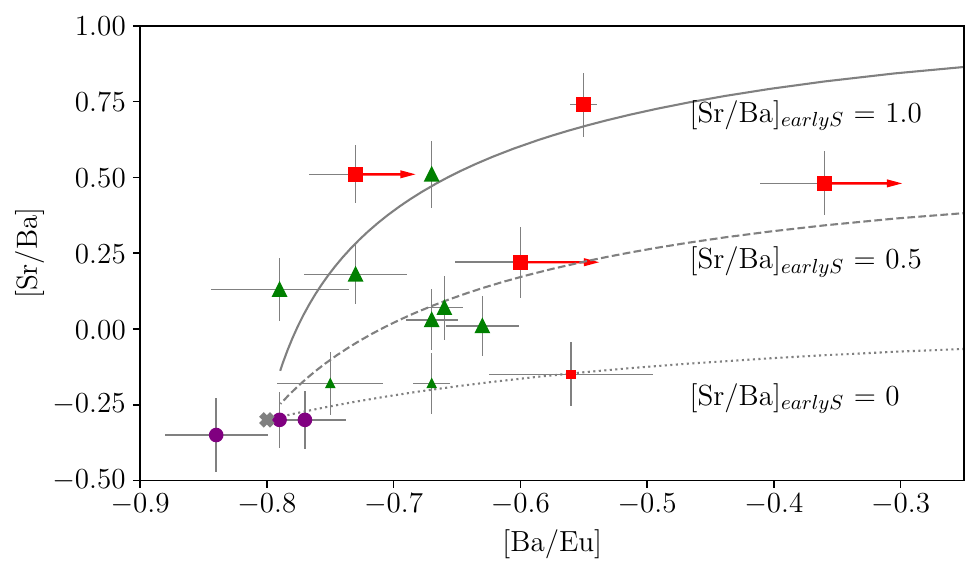}
\caption{Abundance ratios in r-\ii\ (circles), r-\ione\ (triangles) stars, and sample stars with normal [Eu/Fe] $< -0.3$ (squares). Smaller symbols represent stars with [Fe/H] $ > -2.5$. An x-symbol indicates a pure r-process ratios derived from average abundances in r-\ii\ sample stars. Different lines show abundance ratios calculated for mixtures of r-process and s-process with varying [Sr/Ba]$_{\rm earlyS}$. See the legend for designations.}
\label{baeusrba}
\end{figure}

Figure~\ref{baeusrba} shows the [Sr/Ba] ratio as a function of [Ba/Eu] in our sample stars. The [Ba/Eu] ranges from $-0.87$ in an r-\ii\ type star to $-0.36$ in a star with low Eu abundance. Three r-\ii\ sample stars exhibit consistently low [Ba/Eu] values, with an average ratio of [Ba/Eu]$_{\rm r}$ = $-0.81 \pm 0.05$, which aligns with the pure r-process ratio predictions [Ba/Eu]$_{\rm r}$ = $-0.8$ \citep{2014ApJ...787...10B}, $-0.9$ \citep{2020MNRAS.491.1832P},  $-0.7$ \citep{1999ApJ...525..886A}, and an empirical ratio [Ba/Eu]$_{\rm r}$ = $-0.87 \pm 0.06$ based on NLTE analysis of r-\ii\ stars \citep{2014A&A...565A.123M}. 
The [Sr/Ba] ratio spans in a wide range, from [Sr/Ba]$_{\rm r}$ = $-0.31 \pm 0.02$ in r-\ii\ stars to a maximum of 0.74.

\begin{figure}
\centering
\includegraphics[trim={0 5 0 0},width=\hsize]{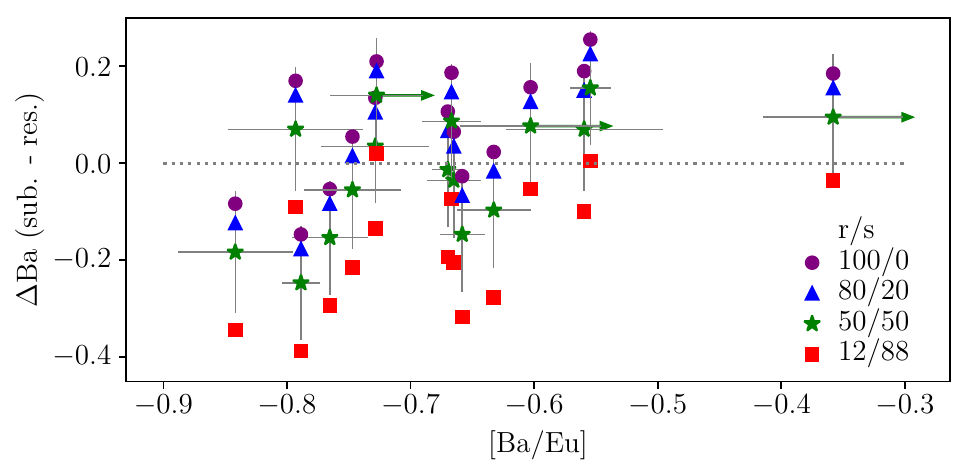}
\includegraphics[trim={0 5 0 0},width=\hsize]{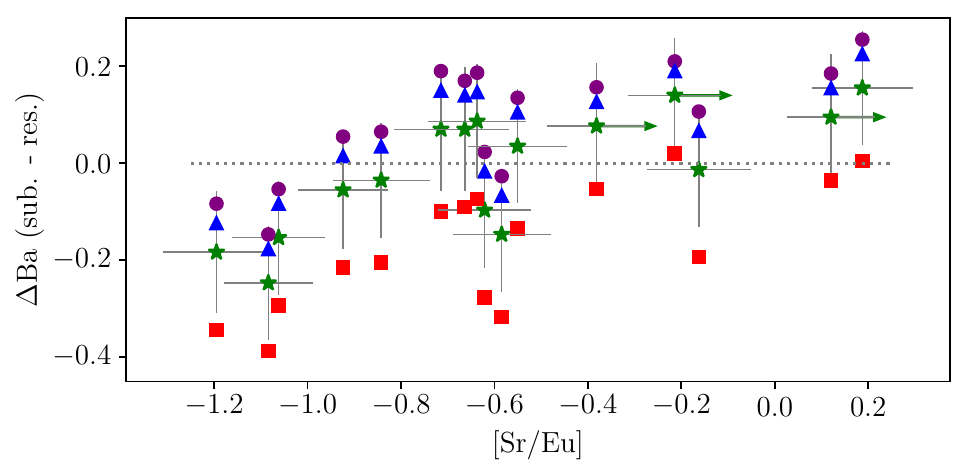}
\includegraphics[trim={0 10 0 0},width=\hsize]{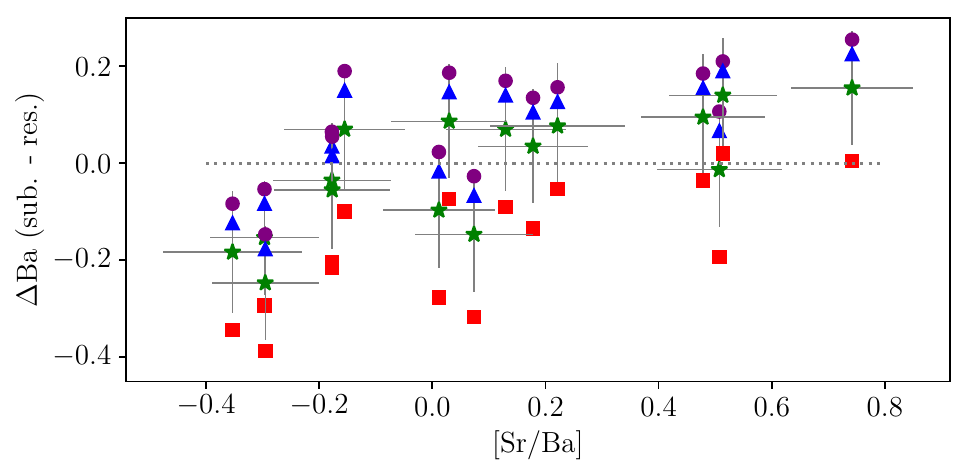}
\caption{Abundance difference between the subordinate and the resonance lines for different r- and s-process fractions (see the legend for designations). The arrows indicate the stars with weak lines of Eu\ii\ with EW $<$ 6~m\AA. 
}
\label{ba_delta_sub_res}
\end{figure}

Figure~\ref{ba_delta_sub_res} presents the abundance difference between the subordinate and resonance lines, $\Delta$Ba~(sub.-res.), computed with different r- to s-process fractions (100\%/0\%, 80\%/20\%, 50\%/50\%, and 12\%/88\%) as a function of abundance ratios. We observe an increasing trend: the fraction of Ba even isotopes rises with [Ba/Eu], suggesting an increasing s-process contribution. It is unsurprising that the s-process produces Ba even isotopes and results in an increasing [Ba/Eu]. The key point is that an increasing trend is also observed when plotting $\Delta$Ba~(sub.-res.) against [Sr/Eu] and [Sr/Ba] (Fig.~\ref{ba_delta_sub_res}, middle and bottom panels). This supports the idea of a common synthesis of additional Sr (beyond that produced via the main r-process) and Ba even isotopes, implying that the extra Sr originates from the s-process. This, in turn, clearly indicates that Sr and Ba are co-produced in this source. We do not claim that this was the only source, however, its contribution to Sr synthesis in the early Galaxy was dominant. Hereafter, we refer to this process as the early s-process.

\subsection{Literature data on Ba isotope measurements}
\subsubsection{Data on individual sample stars}
Applying abundance comparison method in 1D~NLTE for HD~122563, \citet{2008A&A...478..529M} found F$_{\rm odd}$ = 0.22~$\pm$~0.15, which agrees with our measurement of F$_{\rm odd}$ = 0.14$^{+0.09}_{-0.04}$. A slightly higher value of F$_{\rm odd}$ = 0.39~$\pm$~0.06 was obtained by \citet{2019AstL...45..341M} when using modified stellar atmosphere parameters of HD~122563 and an updated Ba~\ii\ model atom. Using the profile fitting method in 1D~LTE, \citet{2012A&A...538A.118G} found a non-physical result, yielding a negative F$_{\rm odd}$~=~$-0.12$~$\pm$~0.07 in  HD~122563.

For HD~4306, \citet{2021A&A...654A.164C} found a good agreement between the observed spectra and 1D~LTE synthetic spectra of the Ba~\ii\ resonance lines when  assuming a pure s-process Ba isotope ratio.

For the r-process enhanced r-\ione\ star HD~108317, \citet{2019AstL...45..341M} found consistent within 0.05~dex NLTE abundances from the resonance and subordinate lines of Ba\ii\ when using F$_{\rm odd}$ = 0.46. Our measurement of F$_{\rm odd}$ = 0.46$^{+0.24}_{-0.15}$ is in agreement with \citet{2019AstL...45..341M}.

For the strongly r-process enhanced r-\ii\ star HE2327--5642, \citet{2010A&A...516A..46M} found a pure r-process Ba isotope ratio in line with our findings.

\subsubsection{Data on MP stars derived from the abundance comparison method}
\citet[][hereafter MZ06]{2006A&A...456..313M} determined NLTE abundances of Eu and Ba, using the r-process and solar Ba isotope mixture for a sample of MW stars. We combined these measurements with Sr NLTE abundance determinations for the same stars presented by \citet{2007ARep...51..903M}. In total, we found 24 MP thick disk and halo dwarfs with Sr, Ba, and Eu abundances along with Ba isotope ratio determinations. The original publications of MZ06 and \citet{2007ARep...51..903M} do not include abundances from individual Ba\ii\ lines. However, we retrieved them from surviving handwritten notes and digitised the data, as presented in Table~\ref{mz06_lbl}.

\begin{table*}
\caption{Stellar parameters and abundances from individual lines of Sr\ii, Ba\ii, and Eu\ii\ from MZ06 and \citet{2007ARep...51..903M}.}
\label{mz06_lbl}
\setlength{\tabcolsep}{0.90mm}
\begin{tabular}{lccccccrrrrrrrrrr}    
\hline
Name & Pop.$^{*}$ &  \teff ,  & log~g &  [Fe/H] & \vt , & [Mg/Fe] & \multicolumn{10}{c}{$\eps$}   \\ 
     &            &    K      &      &         & \small{\kms }  &         & \multicolumn{2}{c}{\small{Ba\ii\ 4554}} & \multicolumn{2}{c}{\small{Ba\ii\ 5853}} & \multicolumn{2}{c}{\small{Ba\ii\ 6496}} & \multicolumn{2}{c}{\small{Sr\ii\ 4215}} & \multicolumn{2}{c}{\small{Eu\ii\ 4129}} \\
\multicolumn{7}{l}{} & \small{r, NLTE} & \small{sol., NLTE} & \small{NLTE} & \small{LTE} & \small{NLTE} & \small{LTE} & \small{NLTE} & \small{LTE} & \small{NLTE} & \small{LTE} \\ 
   \hline 
HD~65583   & t & 5320 & 4.55 & --0.73 & 0.80 & 0.39 & 1.41 &  1.49 &  1.43 &  1.46 &  1.44 &  1.54 &  2.20 &  2.21 &  0.27 &   0.23  \\ 
HD~62301   & t & 5940 & 4.06 & --0.69 & 1.30 & 0.30 & 1.38 &  1.56 &  1.48 &  1.54 &  1.43 &  1.65 &  2.18 &  2.25 &  0.21 &   0.15  \\ 
HD~30649   & t & 5820 & 4.28 & --0.47 & 1.20 & 0.35 & 1.60 &  1.68 &  1.65 &  1.72 &  1.62 &  1.82 &  2.34 &  2.36 &  0.39 &   0.35  \\ 
72~Her     & t & 5735 & 4.24 & --0.34 & 1.00 & 0.38 & 1.71 &  1.83 &  1.74 &  1.81 &  1.74 &  1.94 &  2.52 &  2.54 &  0.54 &   0.51  \\ 
HD~69611   & t & 5820 & 4.18 & --0.60 & 1.20 & 0.43 & 1.39 &  1.55 &  1.52 &  1.57 &  1.48 &  1.69 &  2.35 &  2.37 &  0.30 &   0.26  \\ 
HD~102158  & t & 5760 & 4.24 & --0.46 & 1.10 & 0.40 & 1.63 &  1.70 &  1.63 &  1.70 &  1.61 &  1.83 &  2.44 &  2.46 &  0.42 &   0.38  \\ 
HD~22879   & t & 5870 & 4.27 & --0.86 & 1.20 & 0.44 & 1.24 &  1.39 &  1.34 &  1.37 &  1.31 &  1.48 &  2.13 &  2.17 &  0.10 &   0.04  \\ 
HD~201891  & t & 5940 & 4.24 & --1.05 & 1.20 & 0.41 & 1.05 &  1.20 &  1.10 &  1.10 &  1.11 &  1.21 &  1.85 &  1.89 &--0.09 & --0.16  \\ 
HD~84937   & h & 6350 & 4.03 & --2.07 & 1.70 & 0.36 & 0.12 &  0.19 &       &       &  0.14 &--0.04 &  0.73 &  0.69 &--0.96 & --1.12  \\ 
HD~68017   & t & 5630 & 4.45 & --0.40 & 0.90 & 0.34 & 1.59 &  1.73 &  1.69 &  1.74 &  1.69 &  1.84 &  2.43 &  2.45 &  0.42 &   0.38  \\ 
HD~45282   & h & 5280 & 3.12 & --1.52 & 1.40 & 0.37 & 0.62 &  0.81 &  0.64 &  0.66 &  0.59 &  0.79 &  1.32 &  1.36 &--0.38 & --0.44  \\ 
HD~18757   & t & 5710 & 4.34 & --0.28 & 1.00 & 0.32 & 1.69 &  1.85 &  1.82 &  1.88 &  1.81 &  1.98 &  2.56 &  2.58 &  0.51 &   0.47  \\ 
HD~3795    & t & 5370 & 3.82 & --0.64 & 1.00 & 0.39 & 1.59 &  1.68 &  1.59 &  1.68 &  1.58 &  1.79 &  2.32 &  2.34 &  0.46 &   0.43  \\ 
HD~103095  & h & 5110 & 4.66 & --1.35 & 0.85 & 0.28 & 0.80 &  0.90 &  0.85 &  0.84 &  0.87 &  0.91 &  1.46 &  1.47 &--0.26 & --0.29  \\ 
HD~194598  & h & 6060 & 4.27 & --1.12 & 1.45 & 0.29 & 0.95 &  1.12 &  1.11 &  1.09 &  1.01 &  1.08 &  1.69 &  1.75 &--0.00 & --0.07  \\ 
HD~37124   & t & 5610 & 4.44 & --0.44 & 0.90 & 0.32 & 1.56 &  1.66 &  1.66 &  1.71 &  1.63 &  1.78 &  2.39 &  2.41 &  0.40 &   0.36  \\ 
HD~10519   & t & 5710 & 4.00 & --0.64 & 1.10 & 0.43 & 1.50 &  1.63 &  1.53 &  1.59 &  1.51 &  1.71 &  2.33 &  2.34 &  0.28 &   0.24  \\ 
HD~222794  & t & 5620 & 3.94 & --0.69 & 1.20 & 0.40 & 1.34 &  1.47 &  1.41 &  1.48 &  1.44 &  1.65 &  2.25 &  2.26 &  0.23 &   0.18  \\ 
HD~64606   & t & 5320 & 4.54 & --0.89 & 0.95 & 0.37 & 1.15 &  1.28 &  1.25 &  1.27 &  1.20 &  1.29 &  1.99 &  2.01 &  0.14 &   0.10  \\ 
HD~132142  & t & 5240 & 4.58 & --0.39 & 0.70 & 0.31 & 1.67 &  1.75 &  1.73 &  1.76 &  1.73 &  1.83 &  2.41 &  2.42 &  0.41 &   0.38  \\ 
HD~112758  & t & 5240 & 4.62 & --0.43 & 0.70 & 0.36 & 1.60 &  1.68 &  1.64 &  1.67 &  1.66 &  1.76 &  2.34 &  2.35 &  0.39 &   0.36  \\ 
HD~144579  & t & 5330 & 4.59 & --0.69 & 0.80 & 0.38 & 1.45 &  1.48 &  1.44 &  1.48 &       &       &  2.16 &  2.17 &  0.31 &   0.28  \\ 
HD~148816  & t & 5880 & 4.07 & --0.78 & 1.20 & 0.41 & 1.21 &  1.37 &  1.30 &  1.35 &  1.29 &  1.49 &  2.07 &  2.09 &  0.06 &   0.00  \\ 
HD~193901  & h & 5780 & 4.46 & --1.08 & 0.90 & 0.18 & 1.07 &  1.19 &  1.12 &  1.13 &  1.06 &  1.16 &  1.70 &  1.73 & -0.10 &  -0.16  \\ 
 \hline
\end{tabular}\\
1 -- 't' and 'h' stellar population types correspond to the thick disk and halo, respectively.     \\
Here we present the digitalised hand-written line-by-line abundances used for average abundance calculations in  MZ06 and \citet{2007ARep...51..903M}. We only include MP stars where abundances of the three elements: Sr, Ba, and Eu, are available.  \\
\end{table*}

\begin{figure}
\centering
\includegraphics[trim={0 10 0 0},width=\hsize]{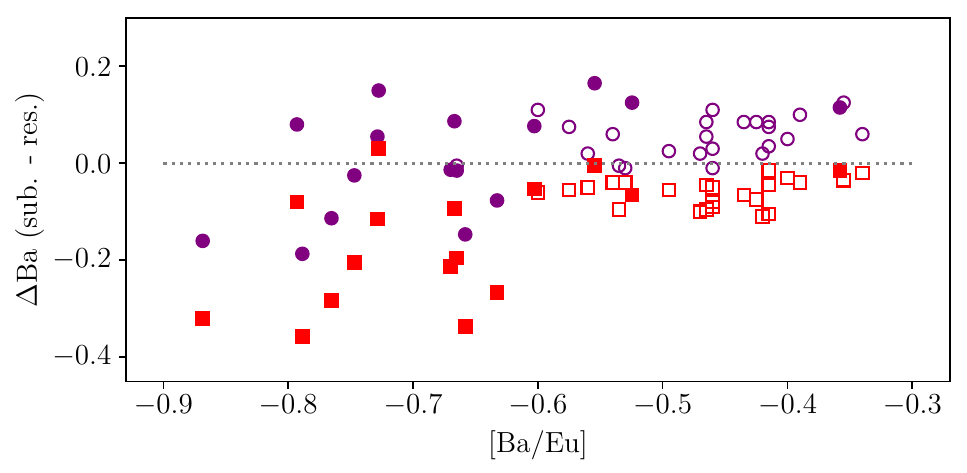}
\includegraphics[trim={0 10 0 0},width=\hsize]{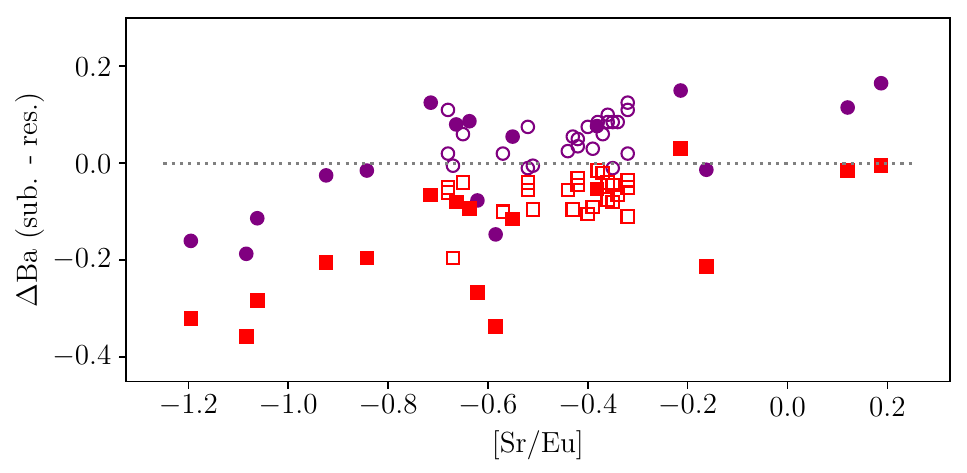}
\caption{Abundance difference between the subordinate and the resonance lines for pure r-process (circles) and solar Ba isotope mixture (squares) in our sample stars (filled symbols) and data from MZ06 (open symbols). For comparison with MZ06, we plot our $\Delta$Ba~(sub.-res.) computed with the r-process Ba isotope mixture from \citet{1999ApJ...525..886A}.}
\label{ba_iso_mz06}
\end{figure}

The majority of the selected MZ06 stars have metallicities in the range $-1.4 <$ [Fe/H] $< -0.3$, with the two stars at [Fe/H] = $-2.1$. The MZ06 sample traces more advanced galactic chemical evolution stages compared to that in the present study and is affected by the main s-process in low- to intermediate-mass AGB stars producing Sr and Ba. However, regardless of the source of Sr and Ba in the MZ06 subsample, the data align well with our findings in the $\Delta$Ba -- [Ba/Eu] and $\Delta$Ba -- [Sr/Eu] diagrams (Fig.~\ref{ba_iso_mz06}). 

\citet{2015A&A...583A..67J} reports a Ba isotope ratio close to the s-process in two VMP stars in Sculptor dSph. These stars exhibit [Sr/Ba] ratios of 0.35 (ET0381) and --0.02 (Scl03059), both higher than the r-process value [Sr/Ba]$_{\rm r}$ = $-0.31$. This finding supports our hypothesis regarding the synthesis of additional Sr via the early s-process.

\subsection{[Sr/Ba] ratio in the early s-prosess}
We aim to determine the chemical properties of the early s-process by subtracting the contribution of the main r-process from the Sr, Ba, and Eu abundances of the sample stars. Here, we assume that the main r-process and the early s-process are the two major sources of n-capture elements in the early Galaxy. Potential additional sources of n-capture elements are discussed in the following section.

The key point is that Sr and Ba originate from both the main r-process and the early s-process, while Eu is exclusively produced by the main r-process. This fact enables using Eu as a reference element.
To estimate the [Sr/Ba]$_{\rm earlyS}$ ratio produced by the early s-process, we use the following data: 
the [Sr/Eu] and [Ba/Eu] ratios in the sample stars; the pure r-process abundance ratios from the r-\ii\ sample stars: [Sr/Eu]$_{\rm r}$ = $-1.1$, [Ba/Eu]$_{\rm r}$ = $-0.8$, and [Sr/Ba]$_{\rm r}$ = $-0.3$; solar abundances, $\eps$(Sr)$_{\rm \odot}$ = 2.88, $\eps$(Ba)$_{\rm \odot}$ = 2.17, $\eps$(Eu)$_{\rm \odot}$ = 0.51 from \citet{2021SSRv..217...44L}. 

In a given star, we can compute the number of Sr particles relative to the number of Eu particles. For example, in the Sun, this value amounts to N(Sr)/N(Eu) =  224. In r-\ii\ stars, the corresponding ratio is N(Sr)/N(Eu) = 18. Now let us compute the ratio in a given Sr-enhanced VMP star with [Sr/Eu] = 0.1, [Sr/Ba] = 0.7, and [Ba/Eu] = --0.6. In this star, [Sr/Eu] = 0.1 corresponds to log~N(Sr)/N(Eu) = [Sr/Eu] + log~N(Sr)/N(Eu)$_{\rm \odot}$ = 2.45, or N(Sr)/N(Eu) = 282. In other words, 18 Sr particles originate from the main r-process, while the remaining 264 Sr particles originate from the early s-process. Applying the same reasoning to the number of Ba particles, we find N(Ba)/N(Eu) = 7 for the pure r-process material and N(Ba)/N(Eu) = 11 in the example star. In this star, 7 Ba particles are produced by the main r-process, while the remaining 4 Ba particles originate from the early s-process. Combining our results for Sr and Ba in the star, we find that, in the early s-process, N(Sr)/N(Ba) = 264/4, or [Sr/Ba] = log~(264/4) -- log~(N(Sr)/N(Ba))$_{\rm \odot}$ = 1.1. 

Thus, we find that the example star with [Sr/Ba] = 0.7 can be represented as a mixture of main r-process material with [Sr/Ba]$_{\rm r}$ = --0.3 and early s-process material with [Sr/Ba]$_{\rm earlyS}$ = 1.1. From these values, we determine that in the example star, the main r-process and the early s-process are mixed in proportions of 75\% and 25\%, respectively.

Now, let us formalise this reasoning and define the procedure for estimating [Sr/Ba]$_{\rm earlyS}$.
In a given star, its Sr, Ba, and Eu abundances can be represented as a mixture of r-process and early s-process material combined in a proportion k$_{\rm mix}$ = N$_{\rm r}$/N$_{\rm earlyS}$, where N$_{\rm r}$ and N$_{\rm earlyS}$ denote the total number of particles produced by the r-process and the early s-process, respectively. Let us define N(Sr)$_{\rm earlyS}$/N(Ba)$_{\rm earlyS}$ as X$_{\rm earlyS}$ and express the number of Sr, Ba, and Eu particles in a given star relative to N(Ba)$_{\rm earlyS}$:\\
N(Sr)$_{\rm *}$ = X$_{\rm earlyS}$ + k$_{\rm mix}\cdot$ N(Sr)$_{\rm r}$/N(Ba)$_{\rm r}$ \\
N(Ba)$_{\rm *}$ = 1 + k$_{\rm mix}$ \\
N(Eu)$_{\rm *}$ = k$_{\rm mix}\cdot$ N(Eu)$_{\rm r}$/N(Ba)$_{\rm r}$ \\
Using the measured abundance ratios in a given star, we deduce:\\ 
k$_{\rm mix}$ = ($\dfrac{\rm N(Ba)_{\rm *}/N(Eu)_{\rm *}}{\rm N(Ba)_{\rm r}/N(Eu)_{\rm r}}$ -- 1)$^{-1}$\\
X$_{\rm earlyS}$ = k$_{\rm mix}\cdot$($\dfrac{\rm N(Sr)_{\rm *}/N(Eu)_{\rm *}}{\rm N(Ba)_{\rm r}/N(Eu)_{\rm r}}$ -- N(Sr)$_{\rm r}$/N(Ba)$_{\rm r}$)\\
And finally, [Sr/Ba]$_{\rm earlyS}$ = log~X$_{\rm earlyS}$ -- ($\eps$(Sr)$_{\rm \odot}$ -- $\eps$(Ba)$_{\rm \odot}$).

To evaluate the effectiveness of [Sr/Ba]$_{\rm earlyS}$ synthesis in the early s-process, we selected sample stars with a significant contribution from this process -- those with [Sr/Ba] $>$ 0.1. From seven stars with $-0.79 <$ [Ba/Eu] $< -0.36$ and 0.13 $<$ [Sr/Ba] $<$ 0.74, we found an average [Sr/Ba]$_{\rm earlyS}$ = 1.02 $\pm$ 0.24. When excluding two stars with weak Eu lines, we obtained [Sr/Ba]$_{\rm earlyS}$ = 1.08 $\pm$ 0.17.

The selected seven stars exhibit different n-capture element abundance ratios, however, they consistently yield high [Sr/Ba]$_{\rm earlyS}$ values. For example, CES1732+2344, with [Sr/Ba] = 0.13 and [Ba/Eu] = --0.79, and CES1402+0941, with [Sr/Ba] = 0.74 and [Ba/Eu] = --0.55, provide close [Sr/Ba]$_{\rm earlyS}$ values of 1.3 and 1.1, respectively. We attribute the variations in n-capture element abundance ratios among these stars to differing relative contributions of the r-process and the early s-process to their chemical composition. For the two stars mentioned above, we estimate that r-process and early s-process material are mixed in the following proportions: 95\%/5\%\ in CES1732+2344 and 55\%/45\%\ in CES1402+0941.

The remaining sample stars are either r-\ione\ or r-\ii\ type stars, dominated by the r-process. An exception is the most metal-rich sample star, CES1436--2906, with [Fe/H] = --2.15. Its chemical composition can be explained by a mixture of r-process and the main s-process, which produces [Sr/Ba] = 0 (Fig.~\ref{baeusrba}).

For stars with moderate [Sr/Ba] enhancement, we found that the additional source of Sr synthesis is characterised by [Sr/Ba]$_{\rm earlyS}$ = 0.5 (Fig.~\ref{baeusrba}). This can be explained either by the main s-process contribution in addition to the early s-process or by variations in the effectiveness of Sr synthesis in the early s-process. 
The latter explanation seems more plausible, as the above stars have metallicities in the range $-3 <$ [Fe/H] $< -2.8$, where the first intermediate-mass stars did not evolve yet to an AGB stage and the contribution of the main s-process is unlikely. A variations in the effectiveness of Sr synthesis in the early s-process are supported by different [Sr/Ba] enhancement observed in different systems: stars in ultra-faint dwarf galaxies exhibit systematically lower [Sr/Ba] ratios compared to those found in classical dSphs \citep{2017A&A...608A..89M,2019ApJ...870...83J,2020A&A...641A.127R,2021MNRAS.504.1183S}.

\subsection{Potential sources of Sr and Ba synthesis beyond the main r-process and early s-process}
When estimating [Sr/Ba]$_{\rm earlyS}$, we assume that the main r-process and the early s-process are the two primary n-capture element sources in the early Galaxy. However, we must consider the potential contribution of the i-process from massive stars \citep{2018ApJ...865..120B} and weak r-process \citep{2009ApJ...692.1517I} to the chemical composition of the sample stars. We do not consider a potential contribution from the $\nu$p process \citep{2022ApJ...929...43G}, as its nucleosynthesis yields for Sr and Ba are not available.

Similar to the r-process, the i-process follows a nucleosynthesis path through neutron-rich nuclei, followed by beta decay. Consequently, like the r-process, the i-process encounters the stable r-only isotopes $^{134}$Xe and $^{136}$Xe, which block the $\beta$-decay path and prevent the formation of $^{134}$Ba and $^{136}$Ba. As a result, the i-process is expected to effectively produce odd Ba isotopes $^{135}$Ba and $^{137}$Ba. For low-mass VMP AGB stars where the i-process occurs in their interiors, \citet{2021A&A...648A.119C} found that $^{137}$Ba is surprisingly more abundant than $^{138}$Ba, which dominates in the solar system, as well as in the main s-process and r-process \citep{1999ApJ...525..886A,2014ApJ...787...10B}. In other words, in terms of Ba isotopes, the contribution from the i-process appears similar to that of the r-process, increasing the fraction of odd Ba isotopes.

\begin{figure}
\centering
\includegraphics[trim={0 10 0 0},width=\hsize]{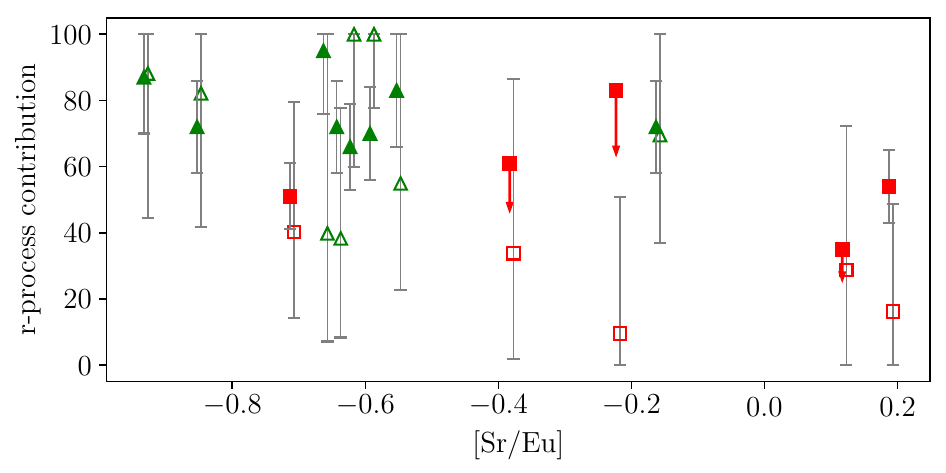}
\caption{R-process contribution to the 13 sample stars (excluding the three r-\ii\ stars) derived from the abundance ratios method (filled symbols) and Ba isotopes method (open symbols) as a function of [Sr/Eu]. Designations are the same as in Fig.~\ref{baeusrba}.}
\label{ba_iso_2methods}
\end{figure}

The r-process contribution to the chemical composition of a given star can be determined through element abundance ratios, as described above, along with another method based on Ba isotope analysis and achieving consistent abundances from the resonance and subordinate lines. For 13 sample stars, excluding the three r-\ii\ type stars, we computed the corresponding r-process fractions that result in $\Delta$Ba~(sub.-res.) = 0. Figure~\ref{ba_iso_2methods} presents a comparison of r-process contributions derived from the two methods. Both methods yield results consistent within the error bars, though the uncertainties are sometimes uncomfortably large.
If there were a significant contribution from the i-process, it would appear as a higher r-process contribution derived from the Ba isotope method compared to that from the element abundance ratio method, which is not observed in our data. The consistency of the two methods supports our hypothesis that, within our detection limits, the r-process and the early s-process are the two major Ba sources in the early Galaxy.

Regarding Sr synthesis, we have demonstrated that, in addition to the main r-process, it is synthesised in the early s-process alongside the Ba even isotopes. However, the weak r-process may also produce a certain amount of Sr \citep{2009ApJ...692.1517I}. How can we identify a possible contribution to Sr abundance from the weak r-process in our sample stars? Let us take another look at Fig.~\ref{baeusrba}. When excluding stars with [Fe/H] $> -2.5$, which may be contaminated by contribution from the main s-process, and r-\ii\ type stars, one can note that the remaining stars can be splitted in two groups according to their predicted [Sr/Ba]$_{\rm earlyS}$: five stars (CES0419--3651, CES0338--2402, CES0215--2554, CES1222+1136, and CES1427--2214) predict [Sr/Ba]$_{\rm earlyS}$ $\simeq$ 0.5, while other five stars (CES1322--1355, CES1402+0941, CES1732+2344, CES0045--0932, CES2232--4138) predict [Sr/Ba]$_{\rm earlyS}$ $\simeq$ 1.0 (see Fig.~\ref{baeusrba}, the dashed and solid line, respectively.) Let us assume that the higher predicted [Sr/Ba]$_{\rm earlyS}$ can be interpreted as significant contamination by Sr produced in the weak r-process in addition to that Sr produced in the early s-process. Therefore, stars from the first group could be considered as having lower (or negligible) contribution from the weak r-process compared to stars from the second group.

Now let us examine the weak r-process yields available in \citet{2009ApJ...692.1517I}. To trace the contribution of the weak r-process, one would need an element produced exclusively by that process. Sr, Y, and Zr are not suitable for distinguishing between contributions from the weak r-process and the early s-process. Figure 12 in \citet{2009ApJ...692.1517I} demonstrates that the weak r-process produces Zn as efficiently as Sr, Y, and Zr. In contrast, the s-process in VMP massive rotating stars produces a negligible amount of Zn \citep[see Fig. 29 in][]{2018ApJS..237...13L}. Therefore, we select Zn as a tracer of the weak r-process. Following our hypothesis regarding the different contributions of the weak r-process to stars in the first and second groups, we might expect them to exhibit distinct Zn abundances, with higher values in the second group stars. Using the Zn abundances determined in LB22, we computed the average [Zn/Fe] = 0.27 $\pm$ 0.08 for the first group and 0.30 $\pm$ 0.12 for the second group. Therefore, we found no significant difference in [Zn/Fe] ratios between the two groups. In both groups, the stars span the same metallicity range, with [Fe/H] from $-2.9$ to $-2.5$, and we found no systematic difference in the other chemical element abundances between the two groups. Based on our results, we found no strong evidence supporting a contribution from the weak r-process  to the chemical composition of our sample stars.

\subsection{Comparison with chemical evolution model predictions}
We employ a stochastic chemical evolution model of the Galactic halo, making use of the \textsc{GEMS} code described in \citet{2025arXiv250320876R} and based on the stochastic models of \citet{2008A&A...481..691C} and \citet{2021MNRAS.502.2495R}. The model traces the early evolutionary stages within the first 1~Gyr of the halo formation, with a timestep of 1~Myr. Stochasticity is introduced by dividing the halo into 10~000 cubic regions, each considered isolated and containing the typical mass of gas swept by a SN-II explosion. Within each region, star formation, gas inflow and outflow are modelled, and gas recycling occurs at each step due to enrichment from evolving stars of different masses.

In the model, we include three sources of n-capture element synthesis: the main r-process, the s-process in rotating massive stars, and the s-process in AGB stars; however, the latter only contributes to the chemical evolution at [Fe/H] $> -2$. We assume that the main r-process occurs in merging neutron stars with progenitor masses of 9-50 M$_{\odot}$. A binary fraction of 0.018 and a fixed coalescence time-scale of 1~Myr are adopted according to \citet{2014MNRAS.438.2177M} and \citet{2015A&A...577A.139C}. We employ the empirical r-process ratios [Sr/Ba]$_{\rm r}$ = $-0.3$ and [Ba/Eu]$_{\rm r}$ = $-0.8$ for consistency with this study. 
The outcome of our chemical evolution results would not change significantly if magneto-rotational SNe are adopted as producers of r-process, instead of merging neutron stars with a very short delay. For the s-process in rotating massive stars (13-120 M$_{\odot}$), we use yields from the \citet{2018ApJS..237...13L} grid, ejected through stellar winds and explosions. For the main s-process in AGB stars, we use the yields from the FRUITY database \citep{2011ApJS..197...17C}. Iron yields are taken from \citet{2018ApJS..237...13L}. For more details on the code, see \citet{2025arXiv250320876R}.

\begin{figure}
\centering
\includegraphics[trim={0 10 0 20},width=\hsize]{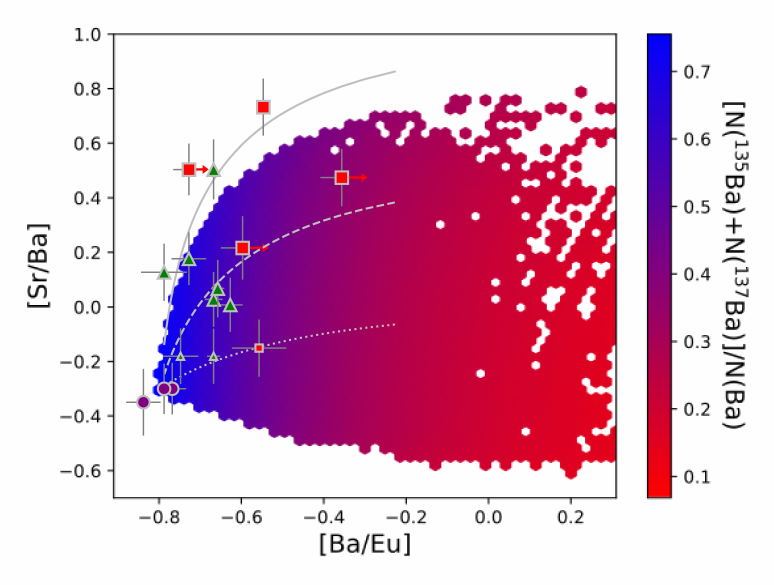}
\caption{Comparison of the derived abundance ratios with the simulated data for long-living stars in the chemical evolution model. Designations are the same as in Fig.~\ref{baeusrba}. The color bar indicates the predicted fraction of odd Ba isotopes.}
\label{baeu_srba_model}
\end{figure}

Figure~\ref{baeu_srba_model} compares the [Sr/Ba] and [Ba/Eu] ratios derived for our sample stars with the simulated data for long-living stars in the chemical evolution model. For consistency, we extracted model data for stars with metallicities in the range $-3.1 <$ [Fe/H] $< -2.1$, matching our sample. The derived abundance ratios generally agree with model predictions. However, reproducing the highest observed [Sr/Ba] values requires a slightly higher predicted [Sr/Ba], which could be achieved by assuming slower rotation velocities in massive stars with $-3.1 <$ [Fe/H] $< -2.1$.

\begin{figure}
\centering
\includegraphics[trim={0 10 0 20},width=\hsize]{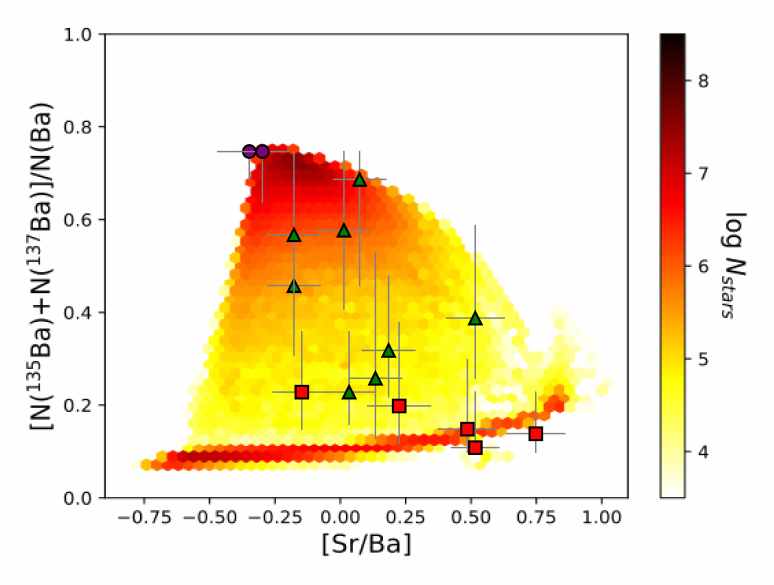}
\caption{Comparison of the derived fraction of odd Ba isotopes with the simulated data for long-living stars in the chemical evolution model. Designations are the same as in Fig.~\ref{baeusrba}. Color bar indicates the predicted number of stars in each area of the diagram. }
\label{isotopes_model}
\end{figure}

Figure~\ref{isotopes_model} compares the derived fraction of odd Ba isotopes as a function of [Sr/Ba] with chemical evolution model predictions. The observed values span a wide range, with F$_{\rm odd}$ varying from pure s-process to pure r-process values. Notably, the model predicts a significant number of stars with [Sr/Ba] $< -0.3$, originating from material produced by massive VMP stars with the highest rotational velocities, as faster rotation leads to lower [Sr/Ba] production. Although our sample does not include stars enriched by such objects, we note that VMP stars with [Sr/Ba] $< -0.3$ should exhibit a low F$_{\rm odd}$ $< 0.2$. Determining Ba isotopic fractions in such stars would provide key constraints on the rotational velocities of massive VMP stars.

\subsection{Other light n-capture elements -- Y and Zr}
\begin{figure}
\centering
\includegraphics[trim={0 10 0 10},width=\hsize]{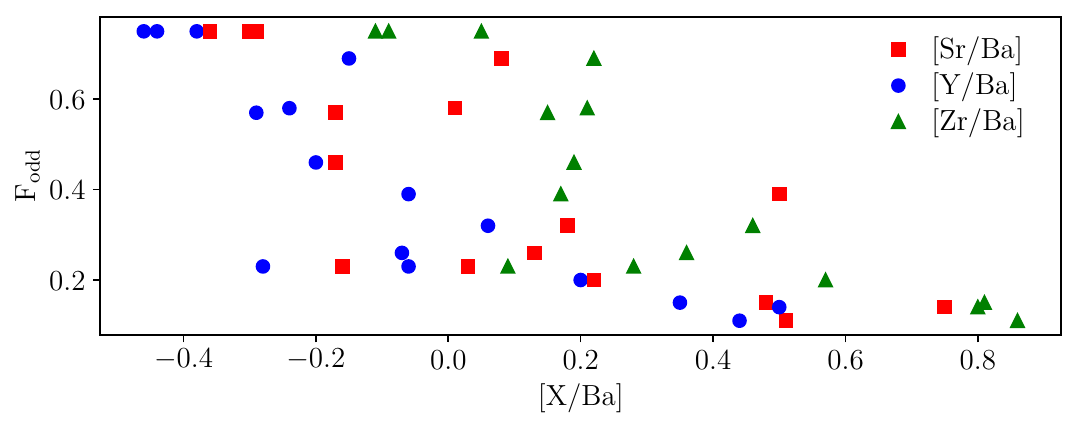}
\caption{F$_{\rm odd}$ as a function of [X/Ba] (see legend) in the sample stars.}
\label{fodd_sryzr}
\end{figure}

We employ Sr as a tracer of light n-capture elements sources, although Y and Zr are also recognised as light n-capture elements and exhibit similar behavior in VMP stars. Our conclusion regarding the joint synthesis of Sr and the even Ba isotopes can also be extended to Y and Zr. Figure~\ref{fodd_sryzr} shows F$_{\rm odd}$ as a function of [Sr/Ba], [Y/Ba], and [Zr/Ba], with Y and Zr abundances taken from LB22. All three elements display a similar trend: F$_{\rm odd}$ decreases with increasing [X/Ba], supporting the idea of co-production of Sr, Y, and Zr alongside the even Ba isotopes.

\section{Conclusions}\label{conclusions}

We present a spectroscopic analysis of 16 VMP stars and determine their Sr, Ba, and Eu NLTE abundances, along with the contributions of the r- and s-processes to their Ba isotope mixture. For Ba isotope ratio analysis, we employ a method based on abundance comparisons between the resonance and subordinate lines of Ba\ii. This method is most effective within a specific stellar atmosphere and barium abundance range, where the EWs of the Ba\ii\ resonance lines range from 60 to 130~m\AA, making them sensitive to the Ba isotope ratio while avoiding excessive saturation. 

Our results provide observational constraints on the source of Sr operating at the early epoch of Galactic chemical evolution. Our findings are summarised as follows:
\begin{itemize}
\item[$\circ$] 
We find a higher s-process contribution to Ba isotopes in stars with greater [Sr/Eu] ratio, suggesting that the additional Sr synthesis was due to the early s-process occurring in massive stars.
Our finding matches the Galactic chemical evolution model predictions of \citet{2013A&A...553A..51C}, \citet{2014A&A...565A..51C},  and \citet{2021MNRAS.502.2495R}.
\item[$\circ$]
The early s-process produces not only Sr but also Ba, therefore it is not a weak s-process. The most likely candidate for the early s-process is a non-standard s-process occurring in massive stars \citep{2012A&A...538L...2F,2016MNRAS.456.1803F,2018A&A...618A.133C,2018ApJS..237...13L}.
\item[$\circ$]
Using Sr-enhanced stars, we estimate that the [Sr/Ba] ratio produced by the early s-process could be as high as [Sr/Ba]$_{\rm earlyS}$ = 1.1 $\pm$ 0.2. 
Regarding the potential synthesis of Sr and Ba in the i-process in massive stars, our results for Ba isotopes and element abundances argues that there was no detectable contribution from this process within the error bars in our sample stars.
\end{itemize}

\section{Data availability}
The full Tables 2 is available at CDS.

\begin{acknowledgements}
The authors thank the referee for carefully reading the manuscript and providing valuable feedback.
We acknowledge A.~Alencastro~Puls for reading the manuscript and providing feedback.
TS is grateful to Yu.~V.~Pakhomov for providing his code for echelle orders merging. 
LL and CJH acknowledge the support by the State of Hesse within the Research Cluster ELEMENTS (Project ID 500/10.006).
FR is a fellow of the Alexander von Humboldt Foundation. 
FR acknowledges support by the Klaus Tschira Foundation.
FR and GC acknowledge the National Recovery and Resilience Plan (NRRP), Mission 4, Component 2, Investment 1.1, Call for tender No. 104 published on 2.2.2022 by the Italian Ministry of University and Research (MUR), funded by the European Union - NextGenerationEU - Project ‘Cosmic POT’ (PI: L. Magrini) Grant No. 2022X4TM3H by the MUR. 
GC thanks for the support INAF (Large Grant 2023, EPOCH and the MiniGrant 2022 Checs).
GC and CJH acknowledge the European Union (ChETEC-INFRA, project no. 101008324).
CJH acknowledges HFHF.
PB acknowledges support from the ERC advanced grant N. 835087 -- SPIAKID.
This study has made use of the Keck Observatory Archive, which is operated by the W. M. Keck Observatory and the NASA Exoplanet Science Institute, under contract with the National Aeronautics and Space Administration. 
 
\end{acknowledgements}

\bibliography{aa55073-25}
\bibliographystyle{aa}

\end{document}